\documentclass[a4paper, 11pt]{article}
\usepackage[raggedright]{titlesec}
\usepackage{amsmath,amssymb,amsfonts}
\usepackage{graphicx,color}
\usepackage{times}
\usepackage{amsmath}
\usepackage{algorithm}
\usepackage{tabularx}

\usepackage[noend]{algpseudocode}
\algtext*{EndIf}
\algdef{SE}[SUBALG]{Indent}{EndIndent}{}{\algorithmicend\ }%
\algtext*{Indent}
\algtext*{EndIndent}

\usepackage{enumitem}
\usepackage{comment}
\usepackage{float}
\usepackage[font={footnotesize}]{caption}
\usepackage{subcaption}
\usepackage{hyperref} 
\usepackage{etoolbox} 
\patchcmd{\thebibliography}{\section*{\refname}}{}{}{}
\usepackage[round]{natbib}
\usepackage{titlesec}
\makeatletter
\newcommand{\multiline}[1]{%
  \begin{tabularx}{\dimexpr\linewidth-\ALG@thistlm}[t]{@{}X@{}}
    #1
  \end{tabularx}
}
\makeatother
\titlespacing{\section}{0pt}{*1.0}{*1.0}
\titlespacing{\subsection}{0pt}{*1.0}{*1.0}

\usepackage{geometry}
 \usepackage{setspace}
\date{}

\usepackage{array,booktabs,arydshln,xcolor}

\usepackage{float}    
\usepackage{verbatim}
\usepackage{listings}
\usepackage{parcolumns}
\begin{document}

\title{\large\textbf{dbnsCavitatingFoam: A density-based solver with equilibrium cavitation models in the OpenFOAM framework}}

\author{\normalsize Mohammad Hossein Arabnejad, Rickard E. Bensow\\
\normalsize \textit{Mechanics and Maritime Sciences, Chalmers University of Technology, 412 96, Gothenburg, Sweden}\\
\normalsize \textit{* Corresponding author. E-mail address: mohammad.h.arabnejad@chalmers.se}\\
}

\maketitle 

\thispagestyle{empty}

\noindent
\small\textit{\textbf{\footnotesize Keywords: } Cavitation; Density-based solver; OpenFOAM; Equilibrium cavitation models} \textit{\footnotesize }
\vspace{0.5cm}

\small
\noindent
\section*{\small Abstract}

This paper presents the development of a density-based  solver suitable for cavitating flows in the OpenFOAM framework. In this solver, the thermodynamic equilibrium mixture approach is adopted to model the presence of and the phase transition between liquid and vapor phases. Using this approach, two cavitation models are implemented in a separate library, although more cavitation models can be easily added. The two are a temperature-dependent cavitation model and a barotropic cavitation model developed by \citet{egerer2014large}. One of the main advantage of the solver is that it considers the compressibilty of all phases. This feature combined with using the density-based  approach enables capturing shock-waves created upon the collapse of cavitating structure which are known to be one of the main mechanism of cavitation erosion. The implementation also includes a post-processing tool which detects aggressive collapse-induced shock-waves based on the method proposed by \citet{mihatsch2015cavitation} and identifies the areas with high risk of cavitation erosion. In order to validate the implemented solver and post-processing tool, four cases with progressively increasing complexity are simulated. The results from simulation of these cases are compared with analytical solution, similar numerical simulations as well as available experimental results. All of  these comparisons show that the numerical results obtained by the implemented solver agrees well with the reference analytical solution and numerical and experimental results.


\section{Introduction}
Hydrodynamic cavitation is defined as the generation of vapor pockets in an accelerating liquid flow due to the pressure drop. It is an unavoidable and mostly detrimental phenomenon leading to a high level of vibration, noise, and erosion in a wide range of hydraulic machines such as hydro-turbines, pumps, diesel injectors and marine propellers \citep{brennen2005fundamentals}. Most of these cavitation related problems occur when the cavities formed in low-pressure regions collapse violently as they are carried into higher pressure region. These collapse events can create strong shock-waves \citep{lauterborn2012cavitation} which are seen as a high level of noise in the far-field and also cause erosion and vibration as they impact nearby surfaces \citep{franc2006fundamentals,tomita1986mechanisms}. In order to avoid these consequences, it is essential to understand the behavior of cavitating flows which are responsible for these cavitation nuisances. This understanding cannot be solely provided by experimental methods as detailed measurements in a cavitating flow is difficult and can hardly be applied on the real-scale hydraulic machine. Besides, these experimental investigations are costly and time-consuming, especially at the early stage of hydraulic machine design.

Considering the limitations of experimental methods, different numerical methods capable of predicting the behavior of cavitating flows have been developed. The most commonly used type of these methods is isothermal transport equation based cavitation model (TEM) where both liquid and vapor are assumed to be incompressible. The cavitation dynamics in these methods are captured by solving the transport equation of the volume fraction of one phase. Although this type of numerical methods has been extensively implemented and used \citep{Ji2015,Arabnejad2018Shedding,bensow2010implicit,asnaghi2015developing}, they have two major drawbacks. First, the transport equations solved for capturing the cavitation dynamics include source terms which take into account the mass transfer between liquid and vapor phases. These source terms typically have several parameters which have been shown to have a strong effect on the captured cavitation dynamics \citep{asnaghi2017improvement}; therefore, they need to be tuned for different flow configurations. Secondly, ignoring the compressibility of the liquid phase in these methods prohibits capturing the collapse-induced shock waves \citep{yakubov2015experience} which are known to be one of the major mechanisms of cavitation nuisances \citep{franc2006fundamentals}.

As an alternative to using incompressible and TEM cavitation models, one can use compressible methods coupled with equilibrium cavitation models. By considering the compressibility effect, these methods are able to capture collapse-induced shockwaves which can then be analyzed to investigate cavitation nuisances such as noise, vibration, and erosion. Furthermore, in contrast to TEM cavitation models, equilibrium cavitation models do not include any tunable parameters  which improves their robustness. Due to these advantages, compressible methods with equilibrium cavitation models have been used to investigate cavitating flows. These investigations mostly include high-speed cavitating flows in diesel injectors or nozzles with the aim to study cavitation erosion. \citet{mihatsch2015cavitation} used these methods to study the cavitating flow in an axisymmetric nozzle. They showed that the location of collapse events leading to strong shockwaves agrees well with the experimental erosion pattern by \citet{franc2011impact}. \citet{orley2017large} studied the cavitating flows in a nine-hole common rail diesel injector using the compressible method with the equilibrium cavitation model. By analyzing the shockwaves captured in the simulations, they were able to examine the hydrodynamics mechanisms responsible for potentially erosive collapse events. In addition to high speed cavitating flow, the above mentioned compressible methods have been used to simulate low-speed cavitating flows around propellers and foils \citep{budich2015numerical,arabnejad2020hydrodynamic}. These simulations were, however, inviscid due to the high computational cost of the compressible methods for low-speed cavitating flows \citep{budichnumerical}.    

Despite the advantages of the above mentioned compressible methods, only a few attempts have been made to implement these methods in open source frameworks \citep{eskilsson2012compressible,deimel2014implementation}. However, the implemented solvers in these attempts have not been made publicly available. In this paper, we present the implementation of a density-based compressible solver with equilibrium cavitation models in the OpenFOAM framework. This implementation follows the programming standard of the OpenFOAM framework; therefore, the implemented solver can serve as a platform for further development of compressible solvers for cavitating flows. Furthermore, the capability of the solver is shown by simulating different cavitating flows and comparing the results with analytical solutions, similar numerical results, and experimental data. This paper is organized into six sections. After this introduction, the theoretical background of the solver is explained in the second section. The implementation of the solver is then explained in the third section and the practical aspects of the solver are presented in the fourth section.  The validation and the capability of the solver are illustrated in the fifth section where the results obtained by the solver are compared with reference analytical, numerical, and experimental results. Lastly, the sixth section presents the conclusions and summary of the paper.

\section{Theoretical background}

\subsection{Governing equations}\label{section:governingEquations}
The implemented solver, which is called \textbf{\texttt{dbnsCavitatingFoam}}, uses the single-fluid approach where the two-phase vapor-liquid mixture is considered as a single compressible medium. Using this approach, the governing equations read as, 

\begin{equation}
\label{Euler_equations}
\frac{\partial  \textbf{U}}{\partial t}+\nabla \cdot \textbf{F}^i = \nabla \cdot \textbf{F}^v,
\end{equation}

where $\textbf{U}= \left \{ \rho , \rho \textbf{u}, \rho E \right\}^T$ represents the vector of conserved variables, $\rho$ is the density of the mixture, $\textbf{u}$ is the velocity vector in the mixture, and $E$ is the total specific energy which is the summation of specific internal energy, $e$, and specific kinetic energy, $\frac{1}{2}\textbf{u} \cdot \textbf{u}$. In the above equations,  $\textbf{F}^i$ and  $\textbf{F}^v$ represent, respectively, the tensor of inviscid and viscous fluxes which can be describe as,
\begin{equation}
\label{FluxesInviscid}
\textbf{F}^i = \left \{ \rho\textbf{u} , \rho \textbf{u}\otimes\textbf{u}+p \mathbf{I}, \rho\mathbf{u}(E+p/\rho) \right\}^T,
\end{equation}
\begin{equation}
\label{FluxesViscous}
\textbf{F}^v = \left \{ 0 ,\tau, \tau\cdot \mathbf{u} - \mathbf{q} \right\}^T,
\end{equation}

where $p$ is pressure, $\mathbf{I}$ is the identity tensor, $\tau$ is the viscous stress tensor, and $\mathbf{q}$ is the heat flux vector. The viscous tensor and heat flux vector can be obtained from,
\begin{equation}
\label{viscoustensor}
\tau=2\mu S - \frac{2}{3}\mu(\nabla\cdot \textbf{u})\textbf{I},
\end{equation}
\begin{equation}
\label{ heatfluxvector}
\textbf{q}= -\kappa \nabla T,
\end{equation}
where  $\mu$ and $\kappa$ are, respectively, the dynamic viscosity and the thermal conductivity of vapor and liquid phases. These two properties are obtained using the homogeneous mixture assumption as,

\begin{equation}
\label{kappaMixture}
\mu = \alpha^v\mu^v+(1-\alpha^v)\mu^l,
\end{equation}

\begin{equation}
\label{kappaMixture}
\kappa = \alpha^v(\frac{\mu^v C^v_{p}}{P_r})+(1-\alpha^v)(\frac{\mu^l C^l_{p}}{P_r}),
\end{equation}
where $\alpha^v$ is the vapor volume fraction, $P_r$ is Prandtl number, $\mu^v$ and $\mu^l$ are, respectively, the dynamic viscosity in vapor and liquid phase, and $C^v_{p}$ and $C^l_{p}$ are, respectively, the specific heat at constant pressure in vapor and liquid phase.

In the \textbf{\texttt{dbnsCavitatingFoam}} solver, equations \ref{Euler_equations} are solved using a density-based approach to obtain the conserved variables, $\rho$, $\rho \textbf{u}$, and $\rho E$ and the equilibrium cavitation model is used to obtain other variables for the mixture. In this type of cavitation models, it is assumed that the two-phase vapor-liquid mixture is in mechanical and thermodynamic equilibrium at every location in the flow and that the transition between liquid and vapor phases is instantaneous. These assumptions imply that the mixture includes pure liquid if the mixture density, $\rho$,  obtained by solving the governing equations, is larger than the liquid saturation density. Similarly, if the mixture density is smaller than the vapor saturation density, $\rho^v_{sat}$, the mixture is assumed to contain only vapor. In case that  $\rho^v_{sat}\leq \rho\leq \rho^{l}_{sat}$, the density is considered as a weighted summation of vapor saturation density and liquid saturation density,  
\begin{equation}
\label{rhoComb}
\rho = \alpha^v\rho^v_{sat}+(1-\alpha^v)\rho^{l}_{sat}.
\end{equation}
For the pure liquid and vapor phases, the vapor fraction is assumed to be, respectively, 0 and 1 while for the mixture of the liquid and vapor, the vapor fraction, $\alpha^v$, is obtained by substituting the mixture density and the saturation densities into equation \ref{rhoComb}. To close the governing equations and obtain the pressure and temperature as a function of density and internal energy, the equation of states included in equilibrium cavitation models are used. Depending on whether these equations of states consider the temperature variation or not, two types of equilibrium cavitation models are implemented in the \textbf{\texttt{dbnsCavitatingFoam}} solver. These implemented models are explained in the following subsections.
\subsubsection*{Temperature-dependent equilibrium cavitation model}
If the effect of temperature variation is taken into account, the cavitation model includes equations of state which gives the pressure and temperature as a function of internal energy and density, $p\left ( \rho,e \right )$, and $T\left ( \rho,e \right )$. For the pure liquid state, ($\rho>\rho^{l}_{sat}$), this equation of state is the modified Tait equation of state \citep{Yuan} which reads as, 

\begin{equation}
\label{TdepTaitEosP}
    p=B\left [ \left (\frac{\rho}{\rho^l_{sat}(T)}  \right )^N-1 \right ]+p_{sat}(T),
\end{equation}
where $B$ and $N$ are the constants and $p_{sat}(T)$ is the saturation pressure. The temperature in the pure liquid is then obtained from a caloric equation of state, 
\begin{equation}
\label{TdepTaitEosT}
e = C^l_{v}(T-T_{ref}) + e^l_{ref},
\end{equation}
where $C^l_{v}$ is the specific heat at constant volume for liquid, $T_{ref}$ is the reference temperature and $e^l_{ref}$ is the internal energy at this reference temperature. For the pure vapor state ($\rho<\rho^v_{sat}$),  the equation of state for calorically perfect gas is then used to obtain the pressure, 
\begin{equation}
\label{TdepPerfectGasEosP}
p=\rho R T
\end{equation}
where $R$ is the specific gas constant. The temperature is then obtained from,
\begin{equation}
\label{TdepPerfectGasEosT}
e = C^v_{v}(T-T_{ref}) + e^l_{ref} +  L^v_{ref},
\end{equation}
where $L_{v,ref}$ is the latent heat of vaporization and $C^v_v$ is the specific heat at constant volume for vapor. In case that the mixture includes both liquid and vapor ($\rho^v_{sat}\leq \rho\leq \rho^{l}_{sat}$),  the pressure is assumed to be equal to the vapor pressure, 
\begin{equation}
\label{Psat}
p=p_{sat}(T),
\end{equation}
while the temperature in the mixture is calculated from,
\begin{equation}
\label{TdepMixtureT}
\rho e = (\rho^v_{sat} \alpha^v C^v_{v}+\rho^v_{sat} (1-\alpha^v) C^l_{v}) (T-T_{ref}) + \rho e^l_{ref} + \alpha \rho^v_{sat} L^v_{ref}.
\end{equation}

It should be mentioned that equations \ref{rhoComb}-\ref{TdepMixtureT} need to be evaluated iteratively as these equations are coupled. In order to determine the states (liquid/vapor/mixture), the density obtained from solving governing equations should be compared with the liquid and vapor saturation densities. These saturation densities, in turn, are a function of temperature which is determined from different equations depending on the state. In order to handle this coupling, equations are evaluated using an iterative algorithm presented in Algorithm \ref{TdepCav}. 
\begin{algorithm}
        \caption{Iterative algorithm in temperature dependent equilibrium cavitation model.}\label{TdepCav}
        \begin{algorithmic}[1]
            \State Obtaining $\rho^{n+1}$ and $e^{n+1}$ in the new time step by solving governing equations
            \State  Setting initial guess for temperature, $T^* = T = T^n$
            \State $m = 0$
            \State $err = 1.0$
            \While {$ err > 10^{-8}$ and $m < 1000$}  
            \State $T^* = 0.5T^*+0.5T$
            \State $m = m + 1$
            \State Obtaining $\rho^v_{sat} (T^*)$, $\rho^l_{sat} (T^*)$, and $p_{sat} (T^*)$
            \If{$\rho^{n+1} \geq \rho^l_{sat}$ }
            \State Obtaining $p$ and $T$ from equations \ref{TdepTaitEosP} and \ref{TdepTaitEosT} and setting $\alpha^v = 0$
            \ElsIf{$\rho^{n+1} \geq \rho^v_{sat}$}
            \State Obtaining $p$, $T$, and $\alpha^v$ from equations \ref{Psat}, \ref{TdepMixtureT}, and \ref{rhoComb}
            \ElsIf{$\rho^{n+1} < \rho^v_{sat}$}
            \State Obtaining $p$ and $T$ from equations \ref{TdepPerfectGasEosP} and \ref{TdepPerfectGasEosT} and setting $\alpha^v = 1$ 
            \EndIf
            \State $err = \left | T-T^* \right | $
            \EndWhile
            \State Updating the pressure, temperature, vapor fraction in the new time step, $p^{n+1} = p$, $T^{n+1} = T$, $\alpha^{v,n+1} = \alpha^v$
        \end{algorithmic}
    \end{algorithm}

\subsubsection*{Barotropic equilibrium cavitation model}
With the barotropic assumption, the pressure is assumed to depend only on density, $p=p(\rho)$, and the temperature becomes decoupled from other variables; therefore there is no need to solve the energy equation. For the pure liquid phase, the pressure is obtained from the Tait equation of state (equation \ref{TdepTaitEosP}) with the assumption that saturated densities are constant. For the mixture phase, the barotropic equation proposed by \citet{egerer2014large} is used. This equation is derived by assuming that the vaporization/condensation process is isentropic and reads as,
\begin{equation}
\label{Egrer EOS}
p=p_{sat} + C \left( \frac{1}{\rho^l_{sat}}- \frac{1}{\rho} \right),
\end{equation}
where $C$ is a model constant.

\subsection{Finite volume approximation}\label{FiniteVolumeApprox}
The governing equations \ref{Euler_equations} are solved in the OpenFOAM framework which uses finite volume approximation on cells with arbitrary cell-shape. The equations in this framework are integrated over the volume of the cell,
$\Omega_i$, 
\begin{equation}
\label{Euler_equations2}
\frac{\partial  \Bar{\textbf{U}}_i}{\partial t}+\frac{1}{\left | \Omega_i \right |}\sum_{j=1}^{N_i}\hat{F}^i\left ( \Bar{\textbf{U}}_L,\Bar{\textbf{U}}_R,\textbf{n}_j \right )\left | S_{ij} \right | = \frac{1}{\left | \Omega_i \right |}\sum_{j=1}^{N_i}\hat{F}^v\left ( \Bar{\textbf{U}} \right )\left | S_{ij} \right |,
\end{equation}
where $\left | \Omega_i \right |$ is the volume of the cell, $N_i$ is number of faces belonging to the cell, $\left | S_{ij} \right | $ is the area of face $j$, $\hat{F}^i$ and $\hat{F}^v$ are, respectively, the approximate inviscid and viscous fluxes. Further, $\Bar{\textbf{U}}_i$ is the volume-averaged conserved variable vector over the cell, $\Bar{\textbf{U}}_L$ and  $\Bar{\textbf{U}}_R$ are the left and right states of face $j$, and $\textbf{n}_j$ is the normal vector of the face $j$. 
The inviscid flux normal to the faces, $\hat{F}\left ( \Bar{\textbf{U}}_L,\Bar{\textbf{U}}_R,\textbf{n}_j \right ) $, are calculated using 
the Mach consistent numerical flux scheme by \citet{schmidt2015low}. In this scheme, the inviscid flux normal to the face $j$ is obtained as,
\begin{equation}
\label{flux}
\hat{F}\left ( \Bar{\textbf{U}}_L,\Bar{\textbf{U}}_R,\textbf{n}_j \right ) = \rho_{L/R}u_{fj} \left \{ 1 ,\textbf{u}_{L/R}, E_{L/R} \right\}^T + \left \{ 0 ,p_{fj} \cdot \textbf{n}_j, p_{fj}u_{fj} \right\}^T ,
\end{equation}
where $u_{fj}$ and $p_{fj}$ are, respectively, the flux speed and pressure at the face $j$ and $\rho_{L/R}$, $\textbf{u}_{L/R}$, and $E_{L/R}$  are, respectively, density, velocity vector, and internal energy in the left or right state depending on the sign $u_{fj}$. If $u_{fj} > 0$, the density, the velocity vector, and the internal energy on the left side of the face is used, otherwise the values on the right side of the face are inserted into equation \ref{flux}. According to \citet{schmidt2015low}, the flux speed, $u_{fj}$, and pressure, $p_{fj}$, are calculated from,
\begin{equation}
\label{flux2}
u_{fj} = \left( \frac{1}{\rho_L+\rho_R} \right )\left ( \rho_L q_L+\rho_R q_R+\frac{p_L-p_R}{c_{fj}}\right ),\; \;
p_{fj} = \frac{p_L+p_R}{2}, 
\end{equation}
where $q_{L/R}$ is the velocity normal to face $j$ and $c_{fj}$ is the speed of sound at the face $j$ which is obtained as,
\begin{equation}
\label{flux}
c_{fj} = max(c_L,c_R,200),
\end{equation}
where $c_L$ and $c_R$ are the speed of sound at the right and left sides of the face. In the pure liquid and pure vapor state, the speed of sound is assumed to be constant and equal to the speed of sound in the saturated liquid and vapor while the speed of sound in the mixture state, $c_m$, is obtained from Wallis formula \cite{Wallis} as,
\begin{equation}
\label{wallis}
\frac{1}{\rho c_m^2}= \frac{\alpha^v}{\rho^v_{sat} c_v^{2}}+\frac{1-\alpha_v}{\rho^l_{sat}c_{l}^{2}},
\end{equation}
where $c_l$ and $c_l$ are, respectively, the speed of sound in the saturated liquid and vapor. In equations \ref{Euler_equations2}-\ref{wallis}, the left and right states, $()_{L/R}$, need to be obtained using the states at the cell centers. Here, these states are obtained by the piece-wise linear reconstruction method as
 \begin{equation}
\label{SecondOrderReconsUleft}
\bar{\textbf{U}}_{L/R}=\bar{\textbf{U}}_{c,L/R}+\Phi_{L/R}\left [ (\nabla \textbf{U})_{c,L/R}(\vec{x}_{j}-\vec{x}_{c,L/R}) \right ]
\end{equation}
where $(\nabla \textbf{U})_{c,L/R}$ is the gradient at the center of the left/right cell, $(\vec{x}_{j}-\vec{x}_{c,L/R})$ is the distance between the center of the left/right cell and the face $j$, and  $\Phi_{L/R} \in \left [ 0,1 \right ]$ is the limiter function.

\section{Implementation in the OpenFOAM framework}

\begin{figure}[ht!]
    \centering
    \includegraphics[width=0.7\linewidth]{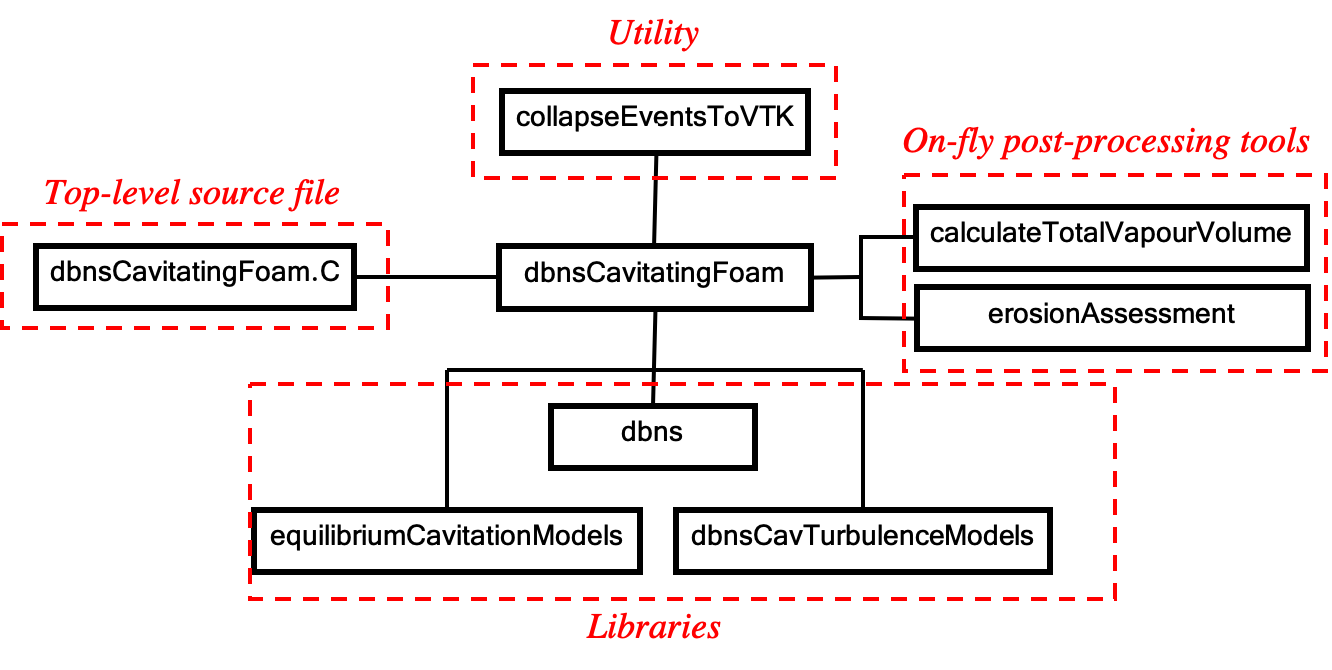}
    \caption{The main components of the implemented solver.}
    \label{fig:implementedSolverComponents}
\end{figure}

The \textbf{\texttt{dbnsCavitatingFoam}} solver is implemented in the OpenFOAM framework \citep{weller1998tensorial} using object-oriented C++. This feature of the solver makes it possible to extend the capability of the solver using the future development in OpenFOAM. The main components of the implemented solver are shown in Figure \ref{fig:implementedSolverComponents}. These components are briefly explained in the following subsections.

\subsection{\texttt{dbnsCavitatingFoam.C} file}
The top-level source file of the solver is \textbf{\texttt{dbnsCavitatingFoam.C}} which includes the implementation of the main algorithm of the solver. This algorithm, shown in Algorithm \ref{Solver_algorithm}, is quite similar to the one implemented in \textbf{\texttt{dbnsFoam}} solver in the \textbf{\texttt{Foam-extend}} package. The algorithm includes a time-marching loop which is performed using a Runge-Kutta method. In each loop of this time marching, the fluxes are first updated using the flux scheme explained in section \ref{FiniteVolumeApprox} and the governing equations are solved explicitly. These equations include the continuity and momentum equations as well as the energy equation if the temperature-dependent cavitation model is used. Afterward, the fields are updated based on the selected cavitation model. At the end of each time marching loop, the turbulent quantities are updated and the selected on-fly post-processing tools are executed.

\begin{algorithm}
        \caption{Algorithm of the solver in \textbf{\texttt{dbnsCavitatingFoam.C}}}\label{Solver_algorithm}
        \begin{algorithmic}[1]
            \While {$t<t_{end}$}   
            \State Runge-Kutta marching loop 
            \Indent
            \State Updating the inviscid fluxes
            \State Solving continuity Eq.
            \State Solving momentum Eq.
            \If{ Cavitation model is temperature dependent}
                \State Solving energy Eq.
            \EndIf
            \State Updating internal and boundary fields based on cavitation model
            \EndIndent
            \State Updating turbulent quantities 
            \State Executing on-fly post-processing tools
            \EndWhile
        \end{algorithmic}
    \end{algorithm}

\subsection{\texttt{equilibriumCavitationModels} library}\label{equilibriumCavitationModels_Library}
\begin{figure}[hb!]
    \centering
    \includegraphics[width=0.5\linewidth]{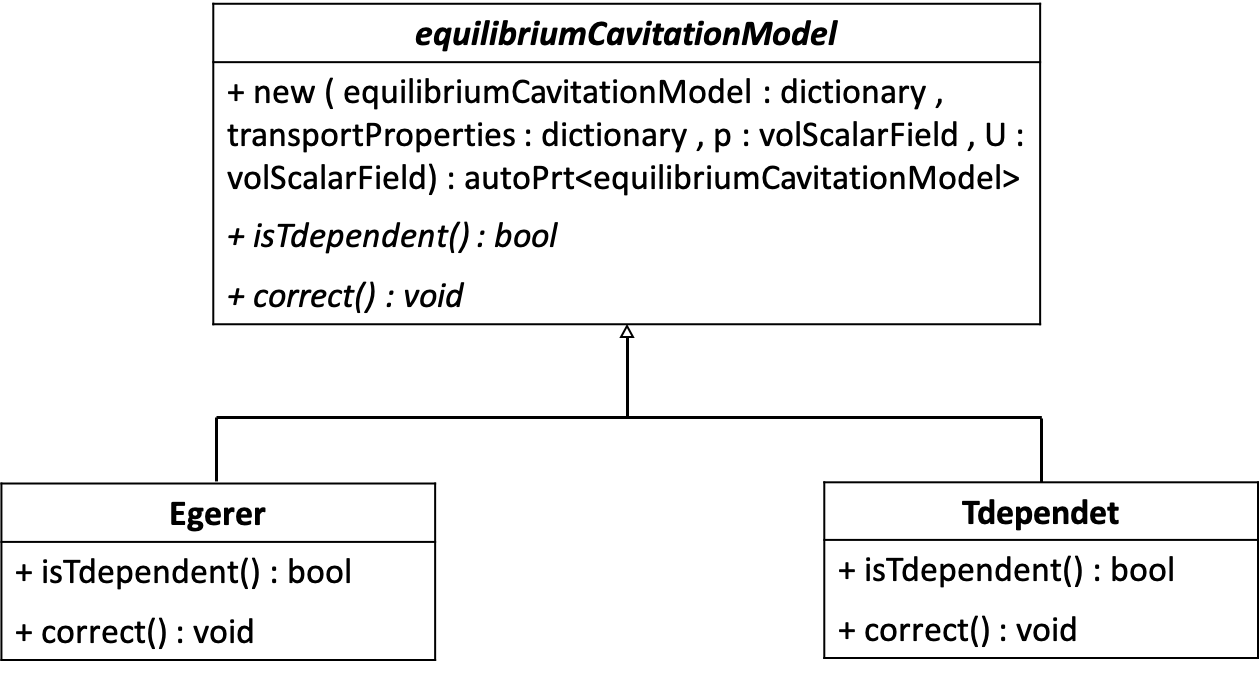}
    \caption{Universal modelling language (UML) diagram of the classes in \textbf{\texttt{equilibriumCavitationModels}} library .}
    \label{fig:equilibriumCavitationModelLib_UML}
\end{figure}

This library includes the implementation of the equilibrium cavitation models described in section \ref{section:governingEquations}. The structure of classes in this library is shown in Figure \ref{fig:equilibriumCavitationModelLib_UML}. The library has an abstract class called \textbf{\texttt{equilibriumCavitationModel}} which is used as the interface of the library with other parts of the solver. This abstract class also includes the implementation of OpenFOAM's run-time selection mechanism which allows the user to select the implemented cavitation models at execution time. One component of this mechanism is a static selector function, \textbf{\texttt{new}} in Figure \ref{fig:equilibriumCavitationModelLib_UML}, which reads the input file, \textbf{\texttt{equilibriumCavitationModelProperties}}, at the execution time and constructs one of the implemented cavitation model based on the input provided in the file. In the current version of the library, two equilibrium cavitation models are implemented as the derived classes of the abstract class, \textbf{\texttt{Egerer}} and \textbf{\texttt{Tdependent}}, although more cavitation models can be easily implemented. These two derived classes include the implementation of two virtual functions declared in the abstract class. The first function is \textbf{\texttt{isTdependent()}} which returns a Boolean indicating whether the cavitation model is temperature-dependent or not. This Boolean is used in the solver to determine if it is needed to solve the energy equation, as shown in the Algorithm \ref{Solver_algorithm}.  The second function is  \textbf{\texttt{correct()}} which updates the internal and boundary values of variable based on the equations presented in section \ref{section:governingEquations}.   

\subsection{\texttt{dbns} library}
The \textbf{\texttt{dbns}} library is a modified version of \textbf{\texttt{dbns}} library in \textbf{\texttt{Foam-extend}} which handles the calculation of the inviscid flux. The implementation of the original dbns library is explained in detail in \citet{ArabnejadDBNS}. Two modifications are made in this library. The first one is to implement the Mach consistent numerical flux scheme which is developed by \citet{schmidt2015low} for cavitating flows. The second  modification is to create an interface to the \textbf{\texttt{equilibiriumCavitationModels}} library in the modified library. To show the necessity of this  modification, Listings \ref{dbns_original} and \ref{dbns_modified} present the constructor of the main class, \textbf{\texttt{numericFlux}}, in the original and modified \textbf{\texttt{dbns}} libraries. As it can be seen, the original \textbf{\texttt{dbns}} library has an interface with the \textbf{\texttt{thermophysicalModels}} library which is used to obtain the thermodynamics variables and speed of sound needed for the flux calculation. In the modified \textbf{\texttt{dbns}}, this interface needs to be replaced with the interface of the \textbf{\texttt{equilibiriumCavitationModels}} library as the implemented solver does not use the \textbf{\texttt{thermophysicalModels}} library. 

\noindent\begin{minipage}{.47\textwidth}
\begin{lstlisting}[language=C++,caption= The constructor of the class \textbf{\texttt{numericFlux}} in the original \textbf{\texttt{dbns}} library,frame=tlrb,  label = dbns_original , captionpos=b]{dbns_original}
numericFlux
(
    const volScalarField& p,
    const volVectorField& U,
    const volScalarField& T,
    basicThermo& thermo
);
\end{lstlisting}
\end{minipage}\hfill
\begin{minipage}{.47\textwidth}
\begin{lstlisting}[caption=The constructor of the class \textbf{\texttt{numericFlux}} in the modified \textbf{\texttt{dbns}} library,frame=tlrb, label =dbns_modified, captionpos=b ]{dbns_modified}
numericFlux
(
    const volScalarField& p,
    const volVectorField& U,
    equilibriumCavitationModel& cavitationModel
);
\end{lstlisting}
\end{minipage}

\subsection{\texttt{dbnsCavTurbulenceModels} library}
This library includes classes which create a new set of turbulence models for the implemented solver. To explain the reason for implementing this library and not using the original sets of compressible turbulence models in OpenFOAM, Listing \ref{MacroTurb_org}  shows one of the macros used to create the compressible turbulence models in OpenFOAM. As it can be seen, this macro needs a reference to the \textbf{\texttt{thermophysicalModels}} library (\textbf{\texttt{fluidThermo}} in Listing \ref{MacroTurb_org}) to create these turbulence models. This reference is used to get access to the transport properties such as viscosity which can be computed as a function of temperature. In the implemented solver, however, these transport properties are computed as a function of vapor volume fraction and are stored in the \textbf{\texttt{equilibiriumCavitationModels}} library; therefore, a new set of turbulence models are created using a reference to the \textbf{\texttt{equilibiriumCavitationModels}} library in the implemented library as it can be seen in the Listing \ref{MacroTurb_mod}. It should be mentioned that the source files of \textbf{\texttt{dbnsCavTurbulenceModels}} library do not include the implementation of individual turbulence models. Instead, this library uses the implementations which are present in the source files of the original turbulence model library in OpenFOAM. Therefore, in order to add a new turbulence model, the implementation of the model should be added to the source files of the original turbulence models library. Furthermore, the  \textbf{\texttt{turbulentDbnsCavModels}} file  in the \textbf{\texttt{dbnsCavTurbulenceModels}} library should be slightly modified so that the new turbulence model is compiled.  

\noindent\begin{minipage}{.47\textwidth}
\begin{lstlisting}[caption= Macro for creating turbulence models for the other compressible solvers in OpenFOAM,frame=tlrb, label =MacroTurb_org, captionpos=b ]{dbns_modified}
makeBaseTurbulenceModel
(
    geometricOneField,
    volScalarField,
    compressibleTurbulenceModel,
    CompressibleTurbulenceModel,
    ThermalDiffusivity,
    fluidThermo
);
\end{lstlisting}
\end{minipage}\hfill
\begin{minipage}{.47\textwidth}
\begin{lstlisting}[language=C++,caption= Macro for creating turbulence models for the implemented solver library,frame=tlrb, label = MacroTurb_mod, captionpos=b ]{dbns_original}
makeBaseTurbulenceModel
(
    geometricOneField,
    volScalarField,
    dbnsCavTurbulenceModel,
    DbnsCavTurbulenceModel,
    ThermalDiffusivity,
    equilibriumCavitationModel
);
\end{lstlisting}
\end{minipage}

\subsection{On-fly post-processing tools}\label{On_fly post_processing_tools}
As shown in Figure \ref{fig:implementedSolverComponents}, the implemented solver has two post-processing tools which may be executed at the end of each time loop. Following the programming standard in OpenFOAM, these two post-processing tools are implemented as function objects. The first tool is \textbf{\texttt{calculateTotalVapourVolume}} which calculates the total volume of the vapor content in the domain and writes out the history of this volume. This history is typically used to obtain the shedding frequency of cavitating structures which is the most common parameter for comparison with the experimental results. The second post-processing tool is \textbf{\texttt{erosionAssessment}} which assesses the risk of cavitation erosion based on the approach presented in \citet{mihatsch2015cavitation}. In this approach, the collapse-induced shock-waves captured by the implemented solver are analyzed to find areas with a high risk of cavitation erosion.  The \textbf{\texttt{erosionAssessment}} provides two outputs for this analysis. The first output is the distribution of the maximum pressure of the faces during the simulation. This distribution identifies areas impacted by strong shock-waves which are assumed to have a high risk of cavitation erosion.  The second output is the location of the collapse events which produce strong shock-waves. The detection of these collapse events is performed using Algorithm \ref{CollapseDetectionAlgorithm}. Compared to the collapse detector algorithm developed by \citet{mihatsch2015cavitation}, the present algorithm has an extra step ( line 3-5 in Algorithm \ref{CollapseDetectionAlgorithm}) in which falsely detected collapse events due to parallel computation are removed. The reason for this error in the collapse detection is that the algorithm includes identifying the neighboring cells of the entries in the \textbf{\texttt{candidateCells}} list (line 2 in Algorithm \ref{CollapseDetectionAlgorithm}). In the OpenFOAM framework, this is performed by calling the \textbf{\texttt{cellCells()}} function in \textbf{\texttt{primitiveMesh}} class. However, this function is not able to detect the neighboring cells across the boundaries between processors in a parallel computation which leads to falsely detected collapse events.  

\begin{algorithm}
        \caption{Procedure of the collapse detection in \textbf{\texttt{erosionAssessment}} functionObject.}\label{CollapseDetectionAlgorithm}
        \begin{algorithmic}[1]
            \State Creating \textbf{\texttt{candidateCells}} list from the cells with the conditions $\alpha^v_{cell} < 0.01$ and $\alpha^v_{cell,old} \geq 0.01$ 
            \State Creating \textbf{\texttt{isolatedCollapseCandidateCells}} list from the cells in \textbf{\texttt{candidateCells}} list which do no have neighboring cells containing vapor 
            \State Removing falsely detected collapse events due to parallel computation 
            \Indent
            \State \multiline{%
            Creating a list from the cells which are filled with water and are neighbor with a cell containing vapor in other processors.}
            \State \multiline{%
            Removing the list of cells created in the previous step from the list of detected collapse events}
            \EndIndent
            \State Recording the pressure at the cells containing collapse events when the sign of $\nabla \cdot \textbf{u}$  changes from negative to positive
            \State  Writing  the location and volume of the cells containing the isolated collapses and the recorded pressure at the center collapse into a file
        \end{algorithmic}
    \end{algorithm}

\subsection{\textbf{\texttt{collapseEventsToVTK}} utility} \label{collapseEventsToVTK}
The \textbf{\texttt{collapseEventsToVTK}} utility reads the output file written by \textbf{\texttt{erosionAssessment}} functionObject and writes this output in VTK format, so that it can be presented using visualization applications, such as ParaView. In order to estimate the aggressiveness of the collapse events in this utility, we follow the work of \citet{schmidt2014assessment} and \citet{mihatsch2015cavitation} which have shown that the maximum pressure captured by Algorithm \ref{CollapseDetectionAlgorithm} is inversely proportional to the size of the cell at the center of the collapse. To remove this mesh dependency, these authors proposed a new parameters called scaled collapse pressure, $p_{scaled}$, as
\begin{equation}
    p_{scaled}=\frac{\sqrt[3]{V_{cell}}}{x_{ref}}p_{collapse},
    \label{p_scaled}
\end{equation}{}
\hspace{-1.0mm}where $p_{collapse}$ is the pressure at the center of the collapse, $x_{ref}$ is the reference length scale, and $V_{cell}$ is the volume of the cell containing the collapse. According to \citet{mihatsch2015cavitation}, $x_{ref}$  is a flow dependent parameter which needs to be estimated if the exact value of scaled pressure is of interest. However, if the aim is to assess and compare the risk of cavitation erosion in similar flow configurations with the same reference length, $p_{scaled} \cdot x_{ref}$ can be used to estimate the aggressiveness of collapse events. This approach is used in this utility. 

\section{Practical aspect of the solver}
The solver is implemented for \textbf{\texttt{OpenFOAM-v1712}}; however, it can be compiled in the newer versions of OpenFOAM with a small modification. The compilation of the solver is done by executing \textbf{\texttt{Allwmake}} which is provided in the main folder of the solver. In addition to this folder, the supplied files include four cases which are used for validation. Similar to other cases in the OpenFOAM framework, the cases have three folders, \textbf{\texttt{0}}, \textbf{\texttt{constant}}, and \textbf{\texttt{system}}. These folders contain essential setting files for running the solver which are explained in detail in \citet{OpenFOAMTutorialGuide}.  Here, we provide only an overview of the setting files which are specific to the implemented solver. 

As mentioned in section \ref{equilibriumCavitationModels_Library}, the \textbf{\texttt{equilibriumCavitationModels}} library includes the implementation of two cavitation models which can be selected by user during execution. For this selection, the library reads the \textbf{\texttt{equilibriumCavitationModelProperties}} file which is placed in the \textbf{\texttt{constant}} folder. An example of the setting in this file where the Egerer cavitation model is selected, is shown in Listing \ref{equilibriumCavitationModelDict}. As can be seen, the entry for the keyword, \textbf{\texttt{equilibriumCavitationModel}} determines which cavitation models to be used. In this file, there is also a subDictionary with the same name as the cavitation model which provides the parameters needed in the selected cavitation model. In the constant folder, there is another file named \textbf{\texttt{tranportProperties}} which is read by the \textbf{\texttt{equilibriumCavitationModels}} library. An example of entries in this file which provide the value for the dynamic viscosity and the heat conductivity for both phases, is shown in Listing \ref{tranportProperties}.

\noindent\begin{minipage}{.47\textwidth}
\begin{lstlisting}[caption= Example of settings provided in \textbf{\texttt{equilibriumCavitationModelProperties}} file ,frame=tlrb, label =equilibriumCavitationModelDict, captionpos=b ]{equilibriumCavitationModelDict}
equilibriumCavitationModel Egerer;
rhoMin          [1 -3 0 0 0] 1;
Egerer
{
    Cl            [0 1 -1 0 0 0 0] 1468.54;
    Cv            [0 1 -1 0 0 0 0] 485.2;
    N             [0 0 0 0 0 0 0] 7.15;
    rhovSat       [1 -3 0 0 0 0 0] 0.01731;
    rholSat       [1 -3 0 0 0 0 0] 998.1618;
    pSat          [1 -1 -2 0 0 0 0] 2340;
    B             [1 -1 -2 0 0 0 0] 3.06e+8;
}
\end{lstlisting}
\end{minipage}\hfill
\begin{minipage}{.47\textwidth}
\begin{lstlisting}[language=C++,caption= Example of settings provided in \textbf{\texttt{tranportProperties}} file,frame=tlrb, label = tranportProperties, captionpos=b ]{dbns_original}
mul     [1 -1 -1 0 0 0 0] 0.001;	
muv     [1 -1 -1 0 0 0 0] 3.57e-7;
cpl     [0 2 -2 -1 0 0 0] 4180.0;;
cpv     [0 2 -2 -1 0 0 0] 1880.0;
Pr      [0 0 0 0 0 0 0] 0.7;
\end{lstlisting}
\end{minipage}

In order to activate the on-fly post-processing tools explained briefly in section \ref{On_fly post_processing_tools}, the code in listing \ref{functionObjects} should be added to the end of the \textbf{\texttt{controlDict}} file in the \textbf{\texttt{system}} folder. For the  \textbf{\texttt{erosionAssessment}} tool, the code includes a keyword called \textbf{\texttt{patches}}, allowing the user to specify the patches on which the erosion assessment is performed. By activating the \textbf{\texttt{erosionAssessment}} tool, the location and the strength of the collapses are written to a file. In the case that simulation is performed in serial mode, the file is located in the main folder of the case, while in the parallel mode, the file is located in processor folders. In both cases, the \textbf{\texttt{collapseEventsToVTK}} utility can be used to read this file and write the collapse events in VTK format as mentioned in section \ref{collapseEventsToVTK}. In parallel mode, however, in order to specify where the utility should look for the file, the parallel option followed by the number of processors should be added to the command of the utility, e.g., \textbf{\texttt{collapseEventsToVTK -parallel 6}}.   
\begin{center}
\noindent\begin{minipage}{.6\textwidth}
\begin{lstlisting}[caption = The code to be added to \textbf{\texttt{controlDict}} file to activate  the on-fly post-processing tools. ,frame=tlrb, label =functionObjects, captionpos=b ]{functionObjects}
functions
{
    calculateTotalVapourVolume
    {
        type    calculateTotalVapourVolume;
        functionObjectLibs ( "libfieldFunctionObjects.so" );
    }
    erosionAssessment
    {
     	type    erosionAssessment;
        functionObjectLibs ( "libfieldFunctionObjects.so" );
        patches ( NozzleLowerWall NozzleUpperWall );
    }
}

\end{lstlisting}
\end{minipage}
\end{center}

\section{Validations}
\subsection{1D cavitating flow}
\begin{figure}[ht!]
\begin{subfigure}{\textwidth}
\centering
\includegraphics[width=0.5\linewidth]{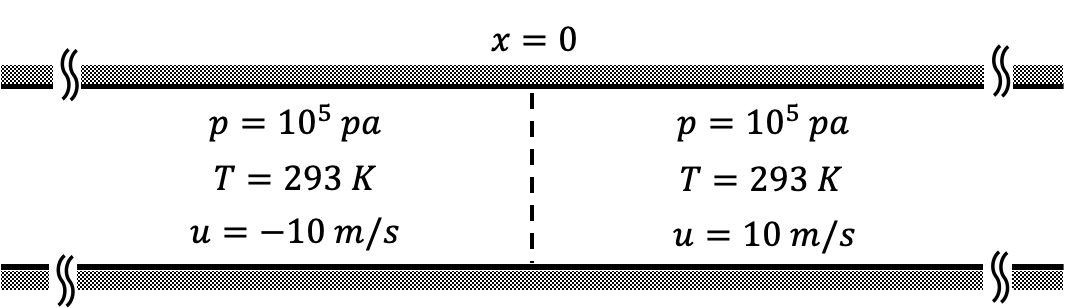}
\caption{}
\label{fig:1DCavConfig}
\end{subfigure}
\begin{subfigure}{0.49\textwidth}
\centering
\includegraphics[width=0.8\linewidth]{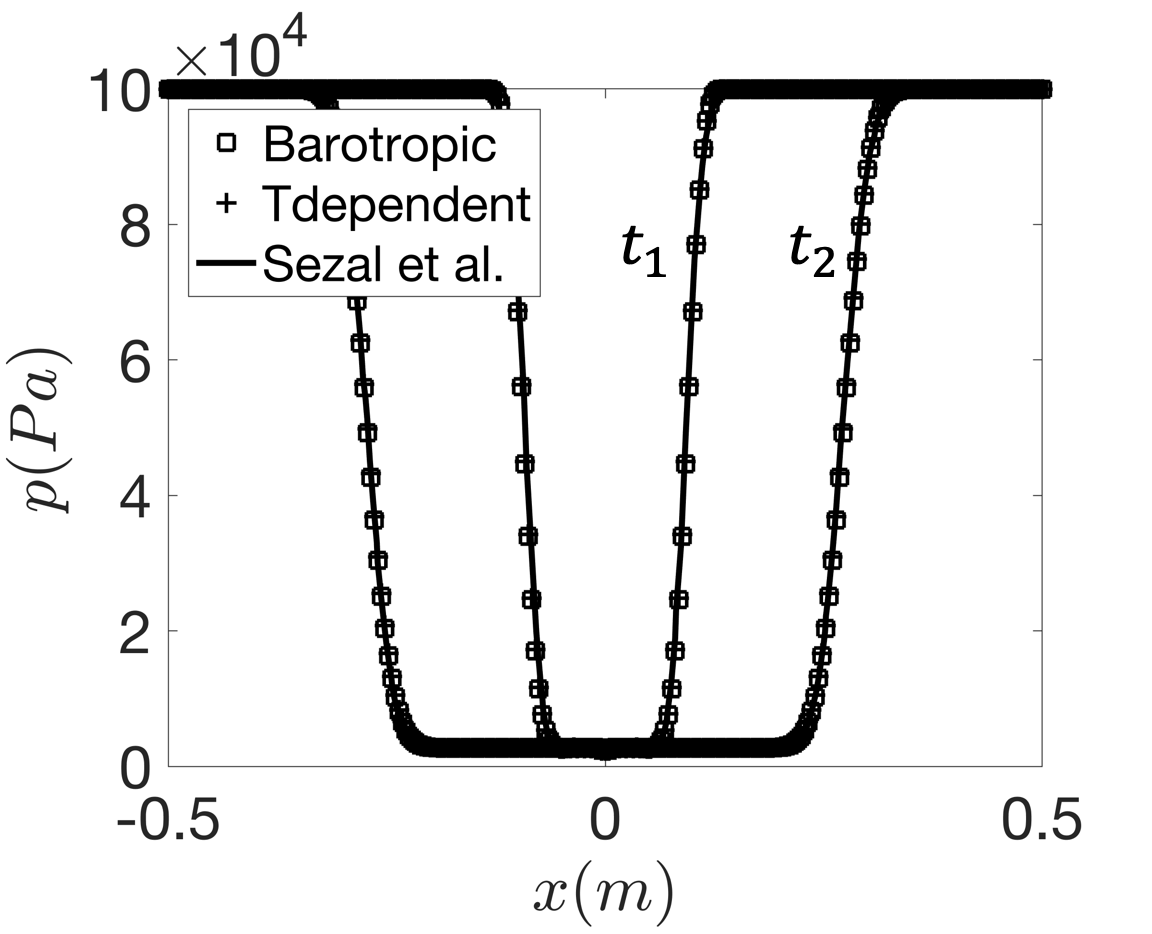}
\caption{}
\label{fig:1DCavP}
\end{subfigure}
\begin{subfigure}{0.49\textwidth}
\centering
\includegraphics[width=0.8\linewidth]{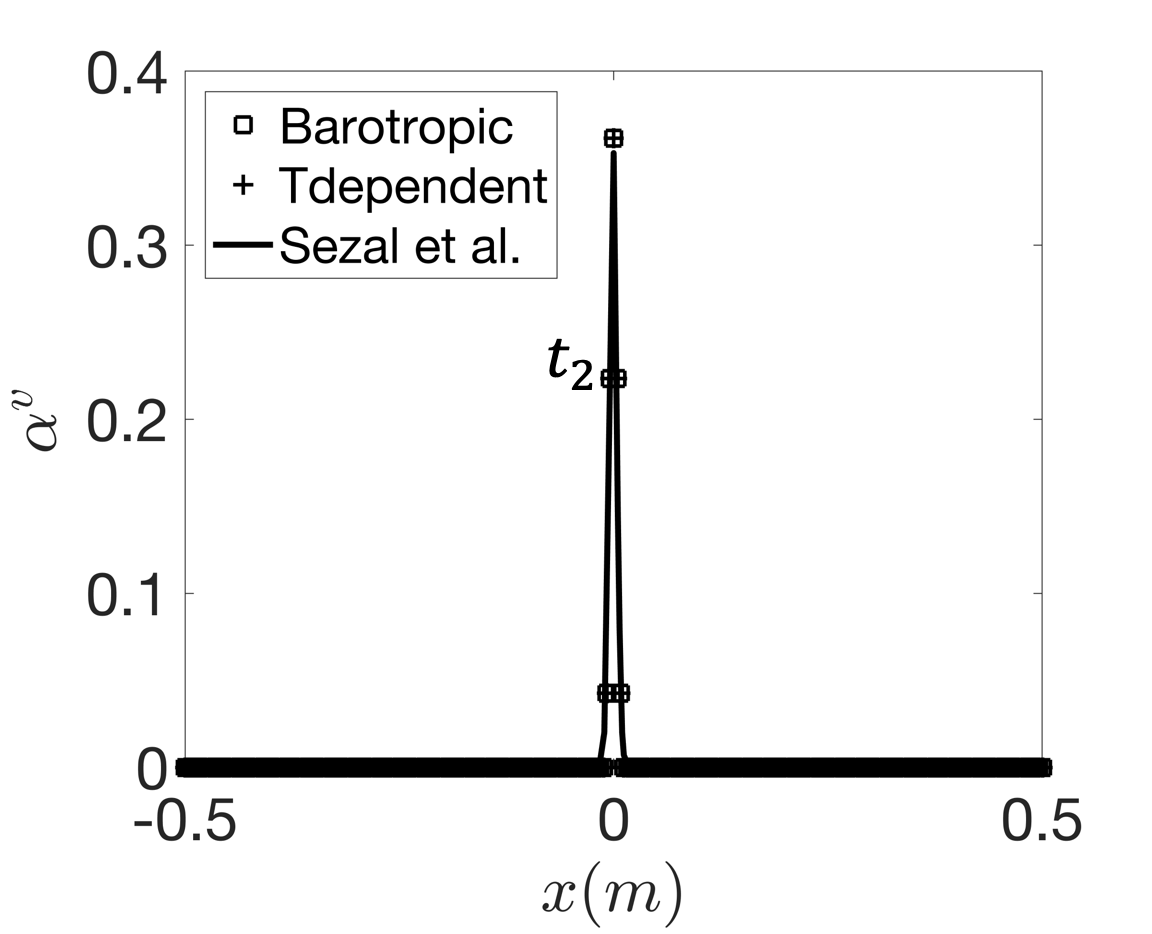}
\caption{}
\label{fig:1DCavAlpha}
\end{subfigure}
\caption{ The simulation of 1D cavitating flow, a) the initial condition, b) comparison between the distribution of the pressure in the simulations with different cavitation models and the simulations by \citet{sezal2009compressible} at $t_1=6\times 10^{-5}s$ and $t_2=1.8\times 10^{-4}s$ , c) comparison between the distribution of the vapor volume fraction in the simulations with different cavitation models and the simulations by \citet{sezal2009compressible} at$t_2=1.8\times 10^{-4}s$.   }
\label{fig:1DCav} 
\end{figure}

The first test case used for validation is a 1D cavitating flow in a tube. This tube has a length of 1 meter and is discretized with 250 cells. Figure \ref{fig:1DCavConfig} shows the initial conditions in this case. At the beginning of the simulation, the tube is filled with water with uniform pressure and temperature, while the initial velocity distribution has a discontinuity at the center of the tube. On the right side and left side of this discontinuity, the velocity is set, respectively, to 10 m/s and -10 m/s. When the simulation starts, the velocity discontinuity creates two expansion waves at the location of the discontinuity,  one of which travels to the right while the other travels to the left. Behind these expansion waves, the pressure drops which can lead to the formation of vapor. 

\newpage

The flow configuration mentioned above does not have an analytical solution due to cavitation formation. However, to validate the implementation of the solver, the numerical results in this paper are compared with the numerical results obtained by \citet{sezal2009compressible}. Figure \ref{fig:1DCavP} shows the pressure distribution along the tube at two time steps, $t_1=6\times 10^{-5}s$ and $t_2=1.8\times 10^{-4}s$ in the simulation using the two implemented cavitation models and the simulation by \citet{sezal2009compressible}. It can be seen that no oscillation around the regions with high gradient pressure is observed. The distribution of vapor fraction in the tube, shown in Figure \ref{fig:1DCavAlpha}, also shows that the amount vapor at the center of the tube formed up to the time step $t_2=1.8\times 10^{-4}s$  is identical in the current simulations and the reference simulations. This agreement for the distributions of pressure and vapor volume fraction demonstrates the validity of the implemented solver and numerical methods.

\subsection{Collapsing bubble}
\begin{figure}[ht!]
\begin{subfigure}{0.69\textwidth}
\includegraphics[width=\linewidth]{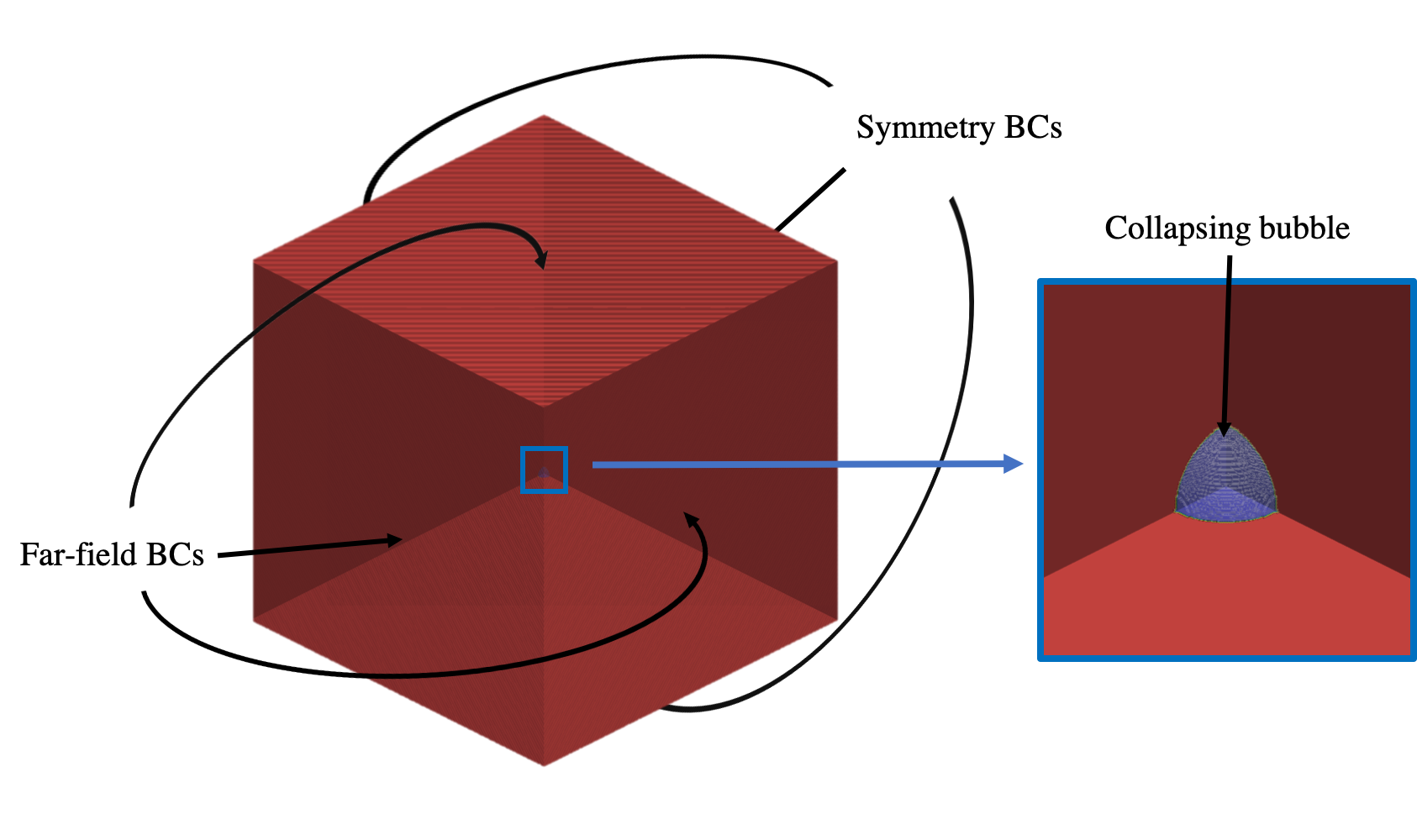}
\vspace{-0.7cm}
\caption{}
\label{3DCollapsingBubble_Config}
\end{subfigure}
\begin{subfigure}{0.3\textwidth}
\vspace{0.9cm}
\includegraphics[width=\linewidth]{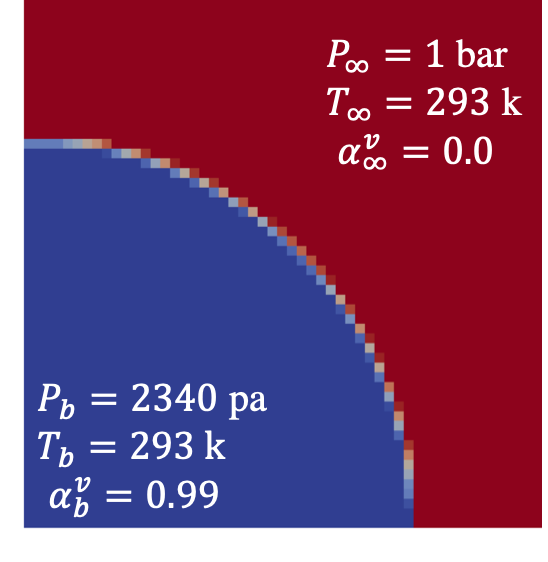}
\caption{}
\label{fig:3DCollapsingBubble_initialCondition}
\end{subfigure}
\caption{3D collapsing  bubble  simulation, a) computational domain and boundary conditions,  b) initial conditions inside and outside of the bubble.}
\label{3DCollapsingBubble_Config_initialCondition} 
\end{figure}

The second test case is the collapse of a spherical bubble in ambient pressure. The computational domain and boundary conditions for the case are shown in Figure \ref{3DCollapsingBubble_Config}. To minimize the computational cost, 1/8 of the bubble with proper symmetry boundary conditions is considered. For far-field boundary conditions, all of the variables are set to a fixed value, except for the velocity for which the zero-gradient boundary condition is used.  The initial radius of the bubble, $R_0$, is $0.4\;\textup{mm}$ which is resolved with 40 cells. The initial conditions for variables are shown in Figure \ref{fig:3DCollapsingBubble_initialCondition}.

\begin{figure}[H]
\begin{subfigure}{0.49\textwidth}
\includegraphics[width=\linewidth]{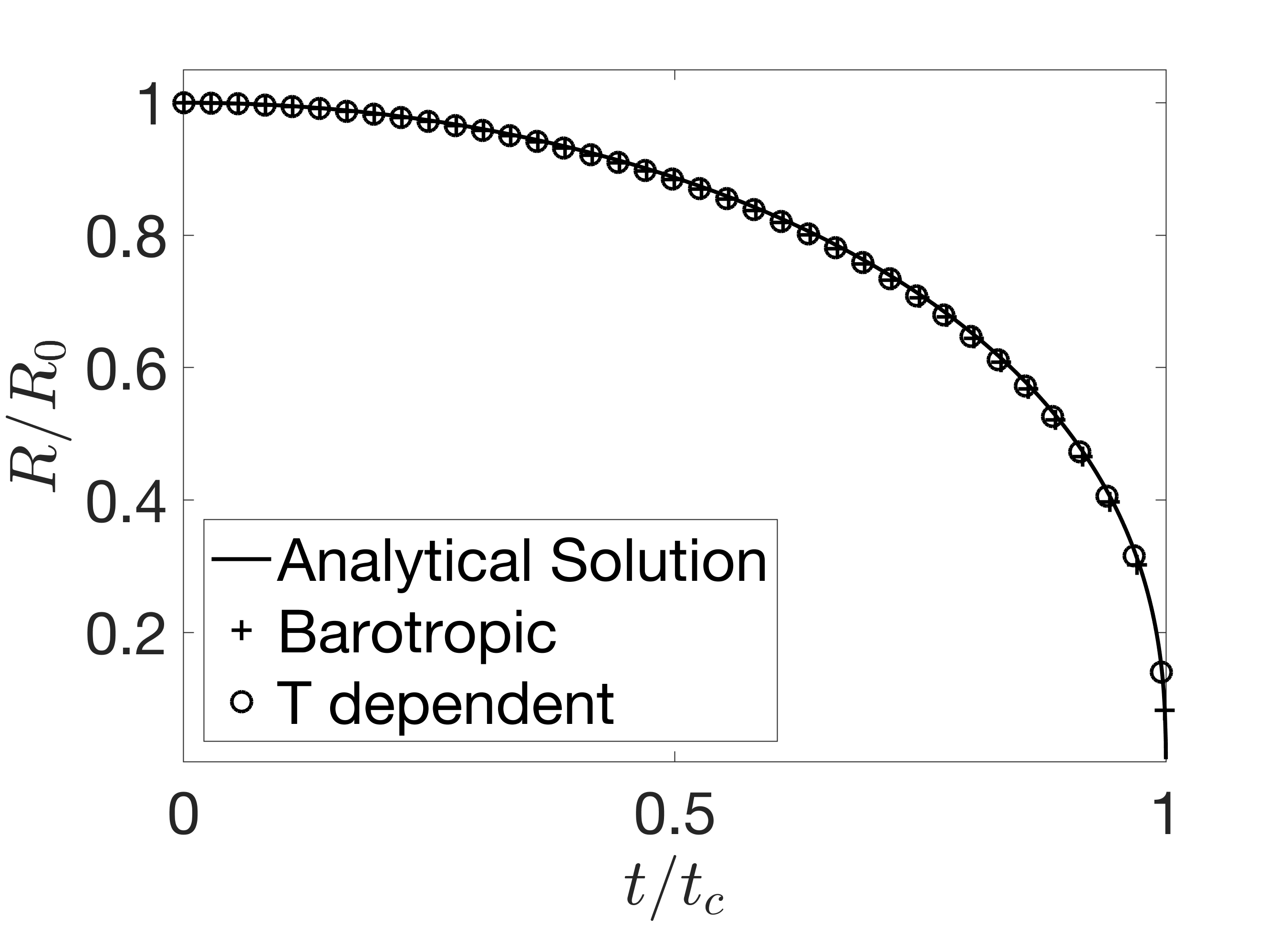}
\caption{}
\label{fig:Bubble_R}
\end{subfigure}
\begin{subfigure}{0.49\textwidth}
\includegraphics[width=\linewidth]{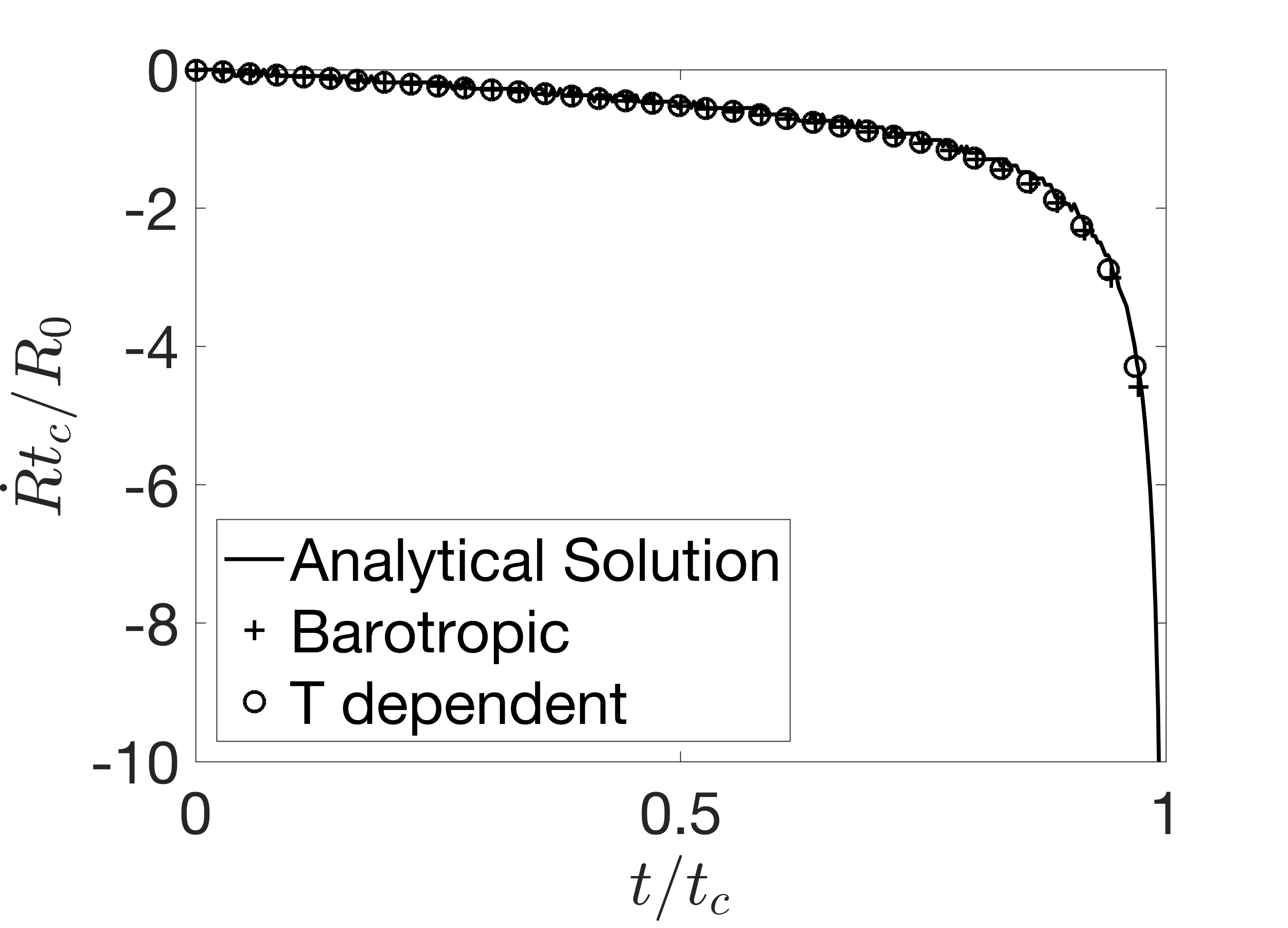}
\caption{}
\label{fig:Bubble_RDot}
\end{subfigure}
\caption{Comparison between the solution of Rayleigh equation and the simulation results obtained by the two implemented cavitation models, a) comparison for non-dimensional radius of bubble, b) comparison for the non-dimensional time derivative of bubble radius. }
\label{fig:Bubble_R_RDot} 
\end{figure}

For the above flow configuration, \cite{rayleigh1917pressure} derived an analytical equation which describes the radius of a spherical bubble collapsing in an incompressible liquid as a function of time. Assuming that the effects of viscosity, the gas inside the bubble, and the surface tension are negligible, this equation can be written as,  

\begin{equation}
R\ddot{R} + \frac{3}{2}\dot{R}^2 = \frac{p_v - p_{\infty}}{\rho_l},
  \label{RaylieghEquation}
\end{equation}

where  $R$ is radius of the bubble, $p_{\infty}$ is the ambient pressure, $p_v$ is the vapor pressure and $\rho_l$ is the density of the surrounding liquid. To validate the implemented solver, the numerical results obtained by the two implemented cavitation models are compared with the solution of equation \ref{RaylieghEquation}.   Figure \ref{fig:Bubble_R_RDot} shows this comparison in terms of the history of bubble radius and time derivative of bubble radius during the collapse. The comparison in this figure indicates that the implemented solver and the cavitation models are valid. It should be mentioned that the comparison between the simulation results and the solution of equation \ref{RaylieghEquation}  is justified, although the compressibility of the surrounding liquid is not taken into account in equation \ref{RaylieghEquation}. According to \citet{gilmore1952growth}, \citet{schnerr2008numerical}, and \citet{sezal2009compressible}, ignoring the compressibility of the surrounding liquid has insignificant effect on the two parameters used for comparison in Figure \ref{fig:Bubble_R_RDot}.

\begin{figure}[ht!]
\begin{subfigure}{0.45\textwidth}
\includegraphics[width=\linewidth]{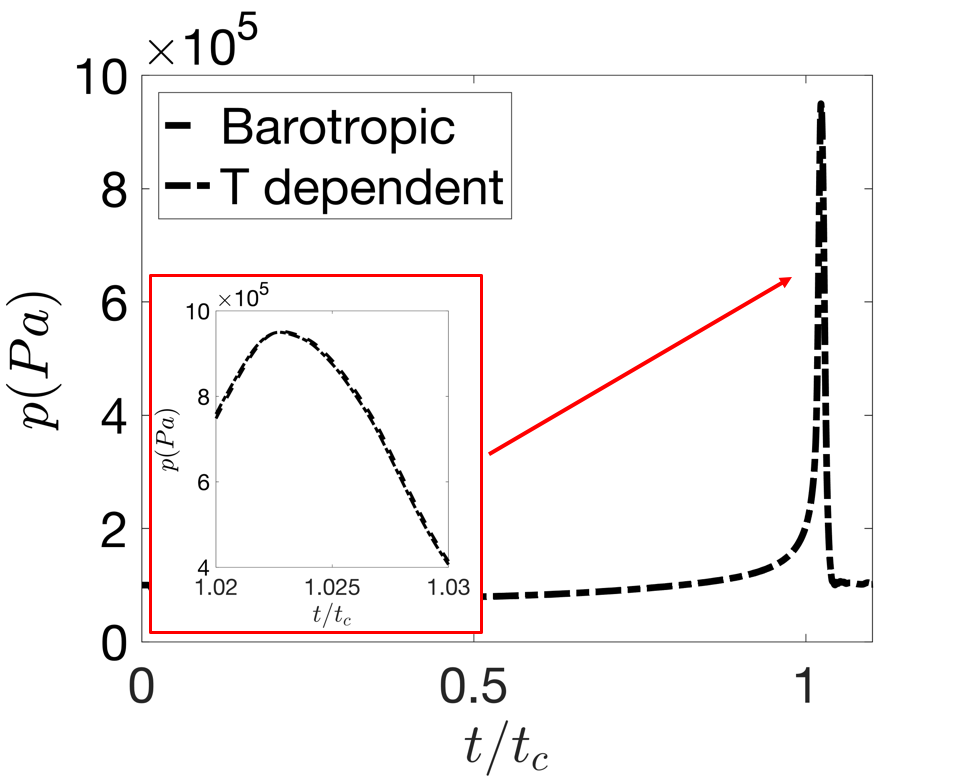}
\caption{}
\label{fig:shockWave}
\end{subfigure}
\begin{subfigure}{0.49\textwidth}
\includegraphics[width=\linewidth]{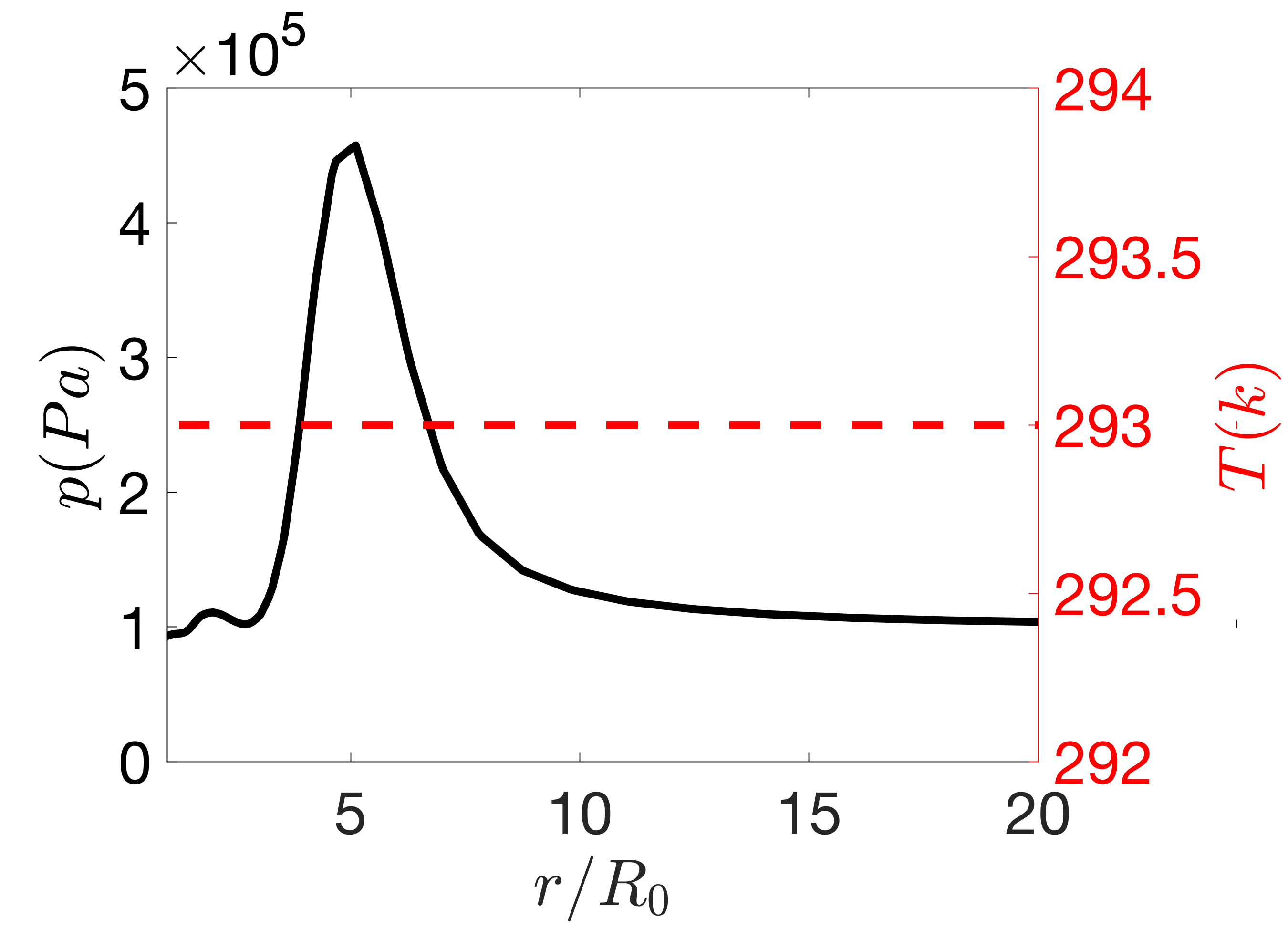}
\caption{}
\label{fig:RadialDist}
\end{subfigure}
\caption{Formation of the shock-wave after the collapse of the cavity, a) Evolution of the pressure probed at $r=2 R_0$ in the simulation with different cavitation models, b) the pressure and the temperature across the collapse-induced shock-wave in the simulation with temperature dependent cavitation model. }
\label{fig:shockWave_RadialDist} 
\end{figure}

As a result of the bubble interface movement toward the center of the collapse (shown in Figure \ref{fig:Bubble_R_RDot}), an inward motion in the surrounding liquid is created during the collapse. This inward motion suddenly stops at the end of the collapse leading to the formation of a shock-wave. To show this shock-wave in the simulations with different cavitation models, Figure \ref{fig:shockWave} shows the history of pressure probed at the radial location of $r=2 R_0$. This history shows a sudden increase and decrease in pressure as the collapse-induced shock-wave passes the location of the probe. The comparison between the magnitude of the pressure increase in the simulation with the barotropic and temperature-dependent cavitation models indicates that the barotropic assumption does not affect the strength of the collapse-induced shock-wave. To provide an explanation for this observation, Figure \ref{fig:RadialDist}
presents the distribution of the pressure and temperature across the collapse-induced shock-wave in the simulation with the temperature-dependent cavitation model. These distributions show that the temperature variation across the shock-wave is insignificant. This uniform temperature distribution indicates that both cavitation models use the same Tait equation of state (equation \ref{TdepTaitEosP}) around the collapse-induced shock-wave which explains the same strength of collapse-induced shock-waves captured by both cavitation models.

\subsection{Cavitating flow in axis-symmetric nozzle}
\begin{figure}[H]
\centering
\includegraphics[width=0.7\linewidth]{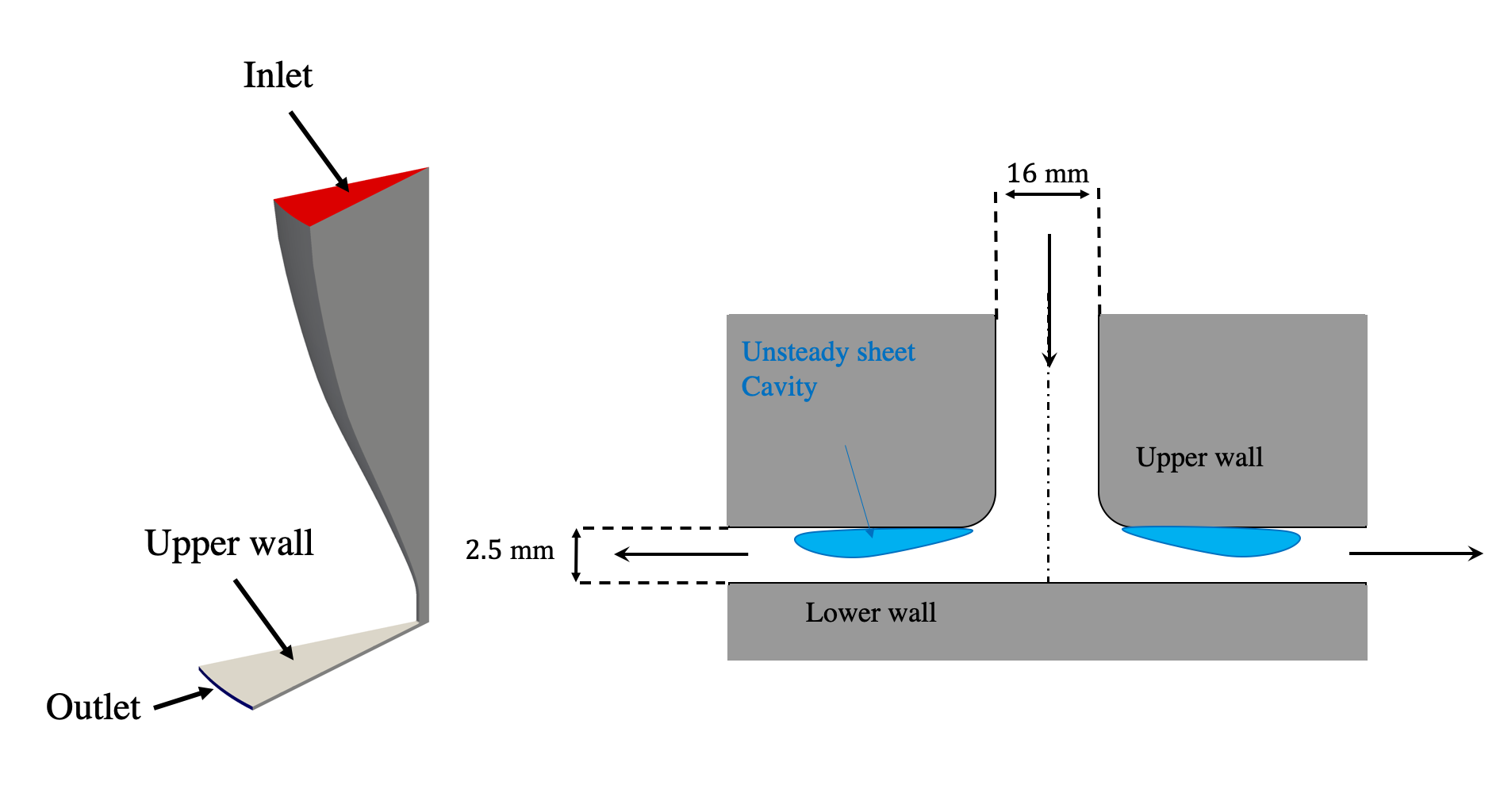}
\caption{Configuration for the axis-symmetric nozzle stagnation flow.}
\label{fig:ComputationalDomain_AxisNozzle} 
\end{figure}

The third case is the inviscid simulation of cavitating flow in an axis-symmetric nozzle which resembles the configuration in the experimental set-up by \citet{franc2011impact} and \citet{gavaises2015visualisation} and the numerical study by \citet{mihatsch2015cavitation}. The flow configuration, shown in Figure \ref{fig:ComputationalDomain_AxisNozzle}, includes a nozzle attached to a disk. The flow enters the nozzle where its velocity increases due to the converging shape of the nozzle. At the exit of the nozzle where the flow is deflected into the disk, the pressure drops which leads to the formation of a sheet cavity attached to the upper wall of the disk. Figure \ref{fig:ComputationalDomain_AxisNozzle} also presents the computational domain used in the simulation which includes only 1/16 of the geometry to minimize the computational cost. The velocity at the inlet boundary and the pressure at the outlet boundary are set to fixed values so that the flow rate and the cavitation number match the ones in the experimental study by \citet{franc2011impact}. 

\begin{figure}[H]
\centering
\includegraphics[width=0.6\linewidth]{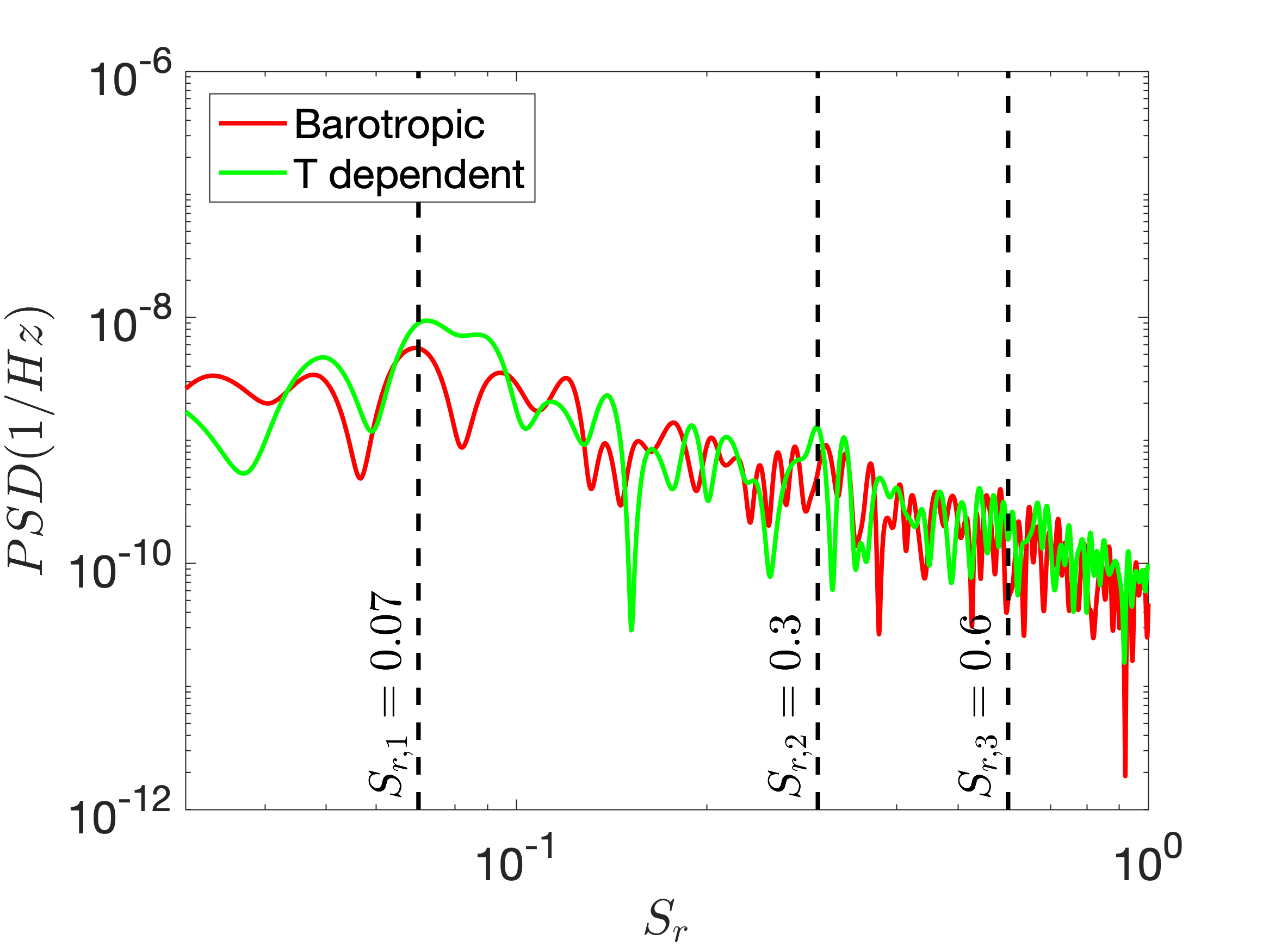}
\caption{Frequency spectra of the total vapor volume signal for the simulations with different cavitation model.}
\label{fig:FrequencyAnalysisAxisNozzle}
\end{figure}

Both previous experimental investigation \citep{gavaises2015visualisation} and numerical simulations \citep{mihatsch2015cavitation} have shown that the sheet cavity attached to the upper wall exhibits a periodic shedding of cavitating structures. The frequency of this shedding can be obtained by applying a fast Fourier transform (FFT) on the total vapor signal obtained by \textbf{\texttt{calculateTotalVapourVolume}} post-processing tool. In order to be able to compare the dominant frequency with the values in similar simulations \citep{mihatsch2015cavitation} and literature, the dominant frequency is reported in term of Strouhal number which is defined as, 

\begin{equation}
S_r = \frac{fL_{c}}{\sqrt{2(p_{in}-p_v)/\rho}}, \end{equation}{}

where $f$ is the frequency, $L_c$ is the maximum length of the sheet cavity, and $p_{in}$ is the inlet pressure. The frequency spectra in the term of the Strouhal number for the simulations with different cavitation models are shown in Figure \ref{fig:FrequencyAnalysisAxisNozzle}. This spectra shows two dominant frequencies, low frequency and high frequency.  The low dominant frequency, which corresponds to the $S_{r,1}=0.07-0.08$, is due to an oscillation of the amount vapor volume in the radial direction according to \citet{mihatsch2015cavitation}. These authors used the same numerics as the current solver coupled with the temperature-dependent cavitation model and found a low-frequency oscillation corresponding to $S_{r,1}=0.07-1.0$.  The high dominant frequency seen in Figure \ref{fig:FrequencyAnalysisAxisNozzle} is due to the shedding of the cavity structure from the attached sheet cavity due to re-entrant jet mechanism as it will be shown later. This dominant frequency corresponds to $S_{r,2}=0.29-0.3$ which is in good agreement the value found by \citet{mihatsch2015cavitation} ,$S_{r,2}=0.27$, and the reported value in the literature (\citet{franc2006fundamentals}), $S_{r,2}=0.25-0.35$. Figure \ref{fig:FrequencyAnalysisAxisNozzle} also shows a local rise in the level of Power Spectral Density (PSD) around the frequencies corresponding to $S_{r,3}=0.6$. This Strouhal number is related to the harmonic of the high 
dominant frequency ($S_{r,3}=2S_{r,2}$).

\begin{figure}[H]
\begin{subfigure}{0.24\textwidth}
\centering
\includegraphics[width=\linewidth]{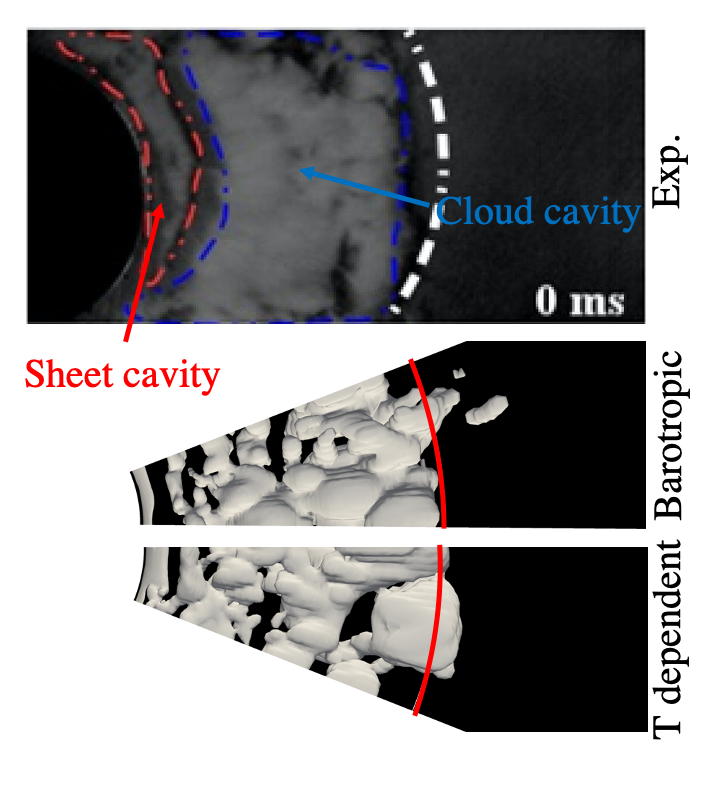}
\caption{$t_1=t$}
\label{Fig:AxisNozzleComshedding1}
\end{subfigure}
\begin{subfigure}{0.24\textwidth}
\centering
\includegraphics[width=\linewidth]{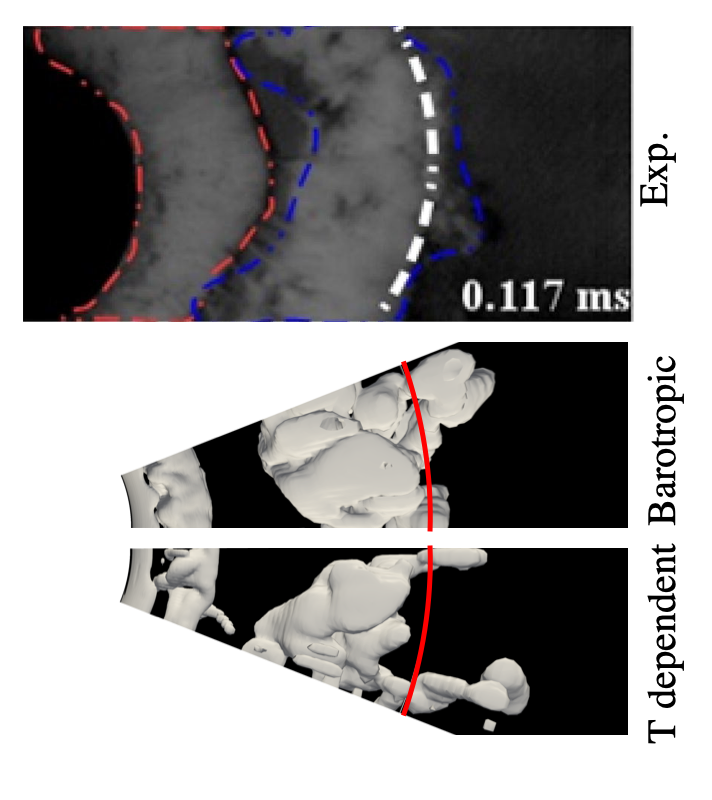}
\caption{$t_2=t+\frac{1}{5}T_s$}
\label{Fig:AxisNozzleComshedding2}
\end{subfigure}
\begin{subfigure}{0.24\textwidth}
\centering
\includegraphics[width=\linewidth]{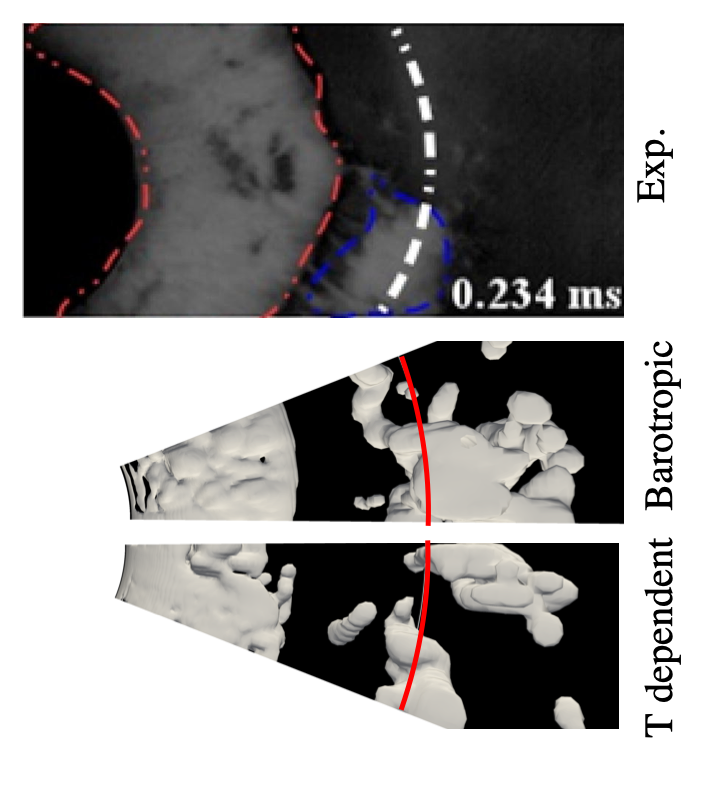}
\caption{$t_3=t+\frac{2}{5}T_s$}
\label{Fig:AxisNozzleComshedding3}
\end{subfigure}
\centering
\begin{subfigure}{0.24\textwidth}
\centering
\includegraphics[width=\linewidth]{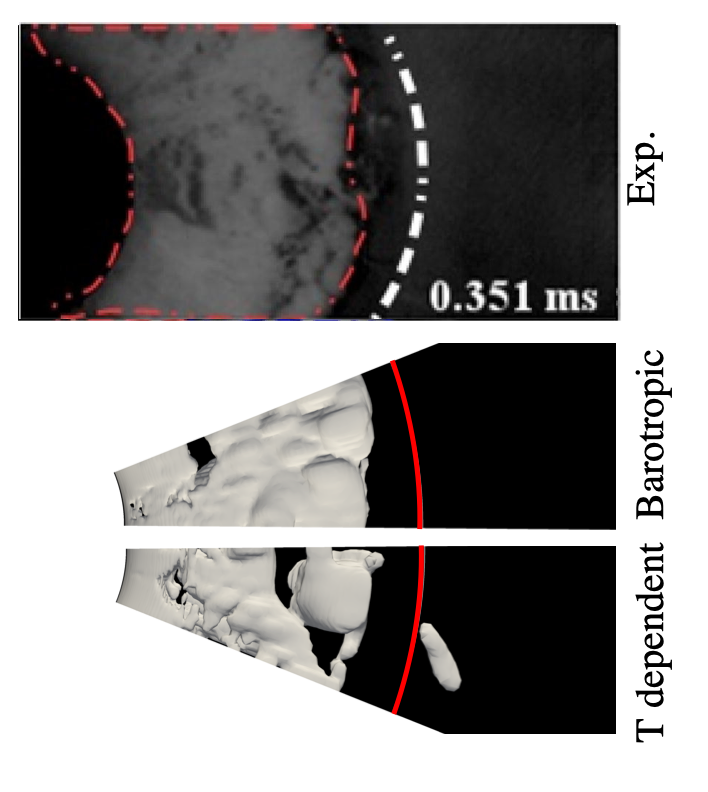}
\caption{$t_4=t+\frac{3}{5}T_s$}
\label{Fig:AxisNozzleComshedding4}
\end{subfigure}
\begin{subfigure}{0.24\textwidth}
\centering
\includegraphics[width=\linewidth]{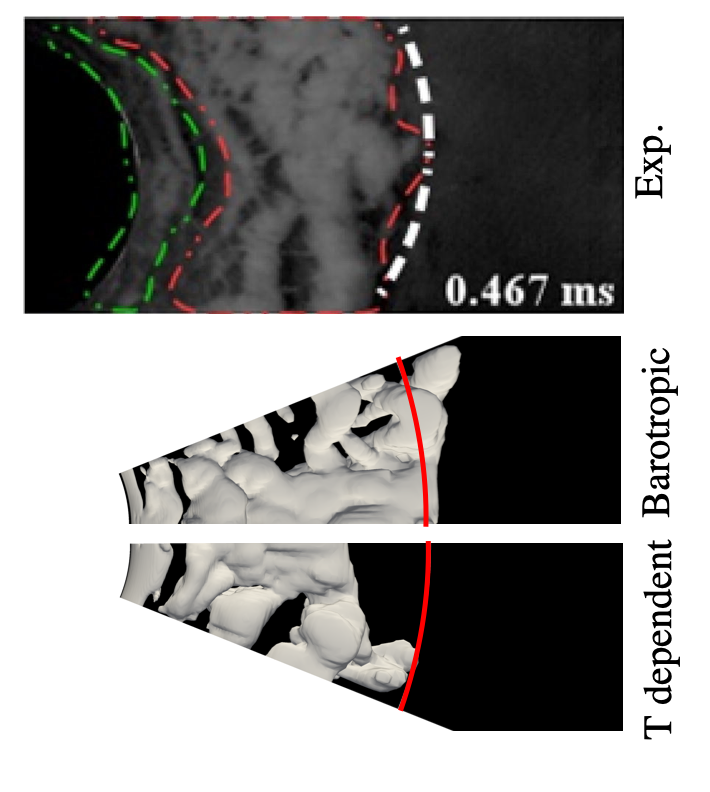}
\caption{$t_5=t+ \frac{4}{5} T_s$}
\label{Fig:AxisNozzleComshedding5}
\end{subfigure}
\begin{subfigure}{0.24\textwidth}
\centering
\includegraphics[width=\linewidth]{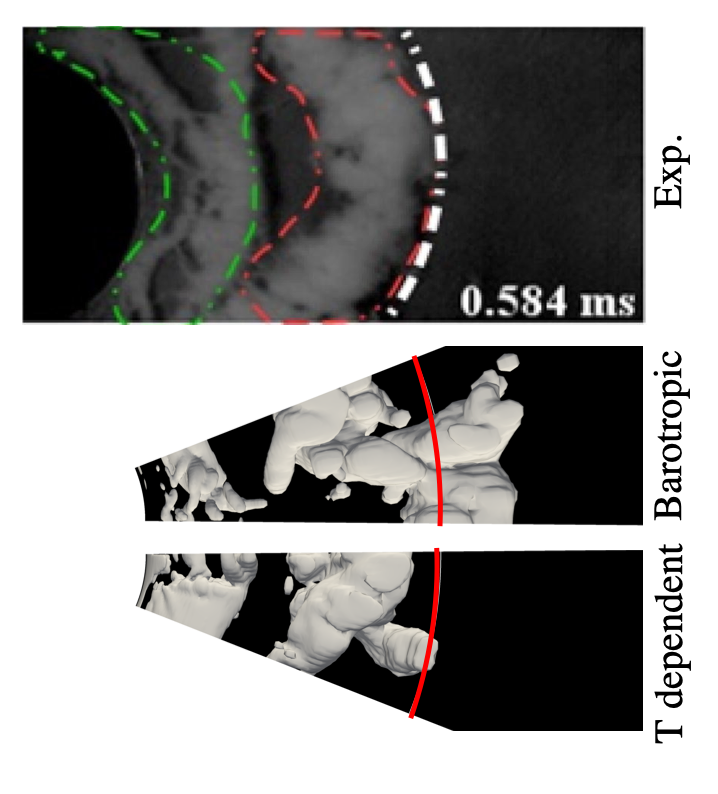}
\caption{$t_6=t+ T_s$}
\label{Fig:AxisNozzleComshedding6}
\end{subfigure}
\caption{Cavitation pattern in one cycle corresponding to the high dominant frequency in the numerical simulations with different cavitation models and the experiment by \citet{gavaises2015visualisation} (The solid red lines in the simulation and dashed white lines in the experiment represent $r=25mm$, $T_s$ and $t$ are, respectively, the high-frequency shedding period and the reference time, and the cavitation pattern in the simulation is shown by iso-surfaces of $\alpha^v= 0.1$)}
\label{Fig:AxisNozzleComCav}
\end{figure}

Figure \ref{Fig:AxisNozzleComCav} shows the periodic shedding of the cavitating structures corresponding to the high dominant frequency ($S_{r,2}$ in Figure \ref{fig:FrequencyAnalysisAxisNozzle}) in the simulations with different cavitation models. For comparison, the figure also includes the experimental images presented in \citet{gavaises2015visualisation} which were obtained at the same cavitation number as the current simulations. The behavior of cavitating structures suggests that the periodic shedding is governed by the re-entrant jet mechanism which has been extensively observed and studied in the cavitating flows over hydrofoils \citep{foeth2008collapse,Callenaere2001,laberteaux2001partial, Arabnejad2018Shedding}. In each period of shedding, a sheet cavity starts to form on the upper wall after the re-entrant jet reaches the entrance of the disk and pinches off a cloud cavity (Figure \ref{Fig:AxisNozzleComshedding1}). This cloud cavity rolls downstream into a high-pressure region where its size decreases due to collapse events occurring in different parts of the cloud cavity. In the meantime, the size of the sheet cavity continuously grows (Figures  \ref{Fig:AxisNozzleComshedding2} and \ref{Fig:AxisNozzleComshedding3}).  As the cloud cavity travels further downstream, a large-scale collapse event happens in the cloud leading to its complete disappearance. At this instance, the sheet cavity has grown to its maximum length and a re-entrant jet has formed at its closure line (Figure  \ref{Fig:AxisNozzleComshedding4}). This re-entrant jet travels upstream and reaches the entrance where it pinches of the cloud cavity (Figure  \ref{Fig:AxisNozzleComshedding5}). After this pinch-off, a new sheet cavity forms and the cycle repeats itself (Figure  \ref{Fig:AxisNozzleComshedding6}). The comparison between numerical results and experimental images indicates that the implemented solver can capture the main behavior of the cavitating structures in the experiment. In the simulation with both cavitation models, the sheet cavity grows to a size which is comparable to the size of the sheet cavity in the experimental images (Figure  \ref{Fig:AxisNozzleComshedding4}).  Further, the collapse location of the cloud cavity in the simulations agrees well with this location in the experimental images (Figure  \ref{Fig:AxisNozzleComshedding4}). Capturing the correct location of this collapse event is important as it is associated with a high risk of cavitation erosion. 

\begin{figure}[H]
\begin{subfigure}{0.39\textwidth}
\centering
\includegraphics[width=\linewidth]{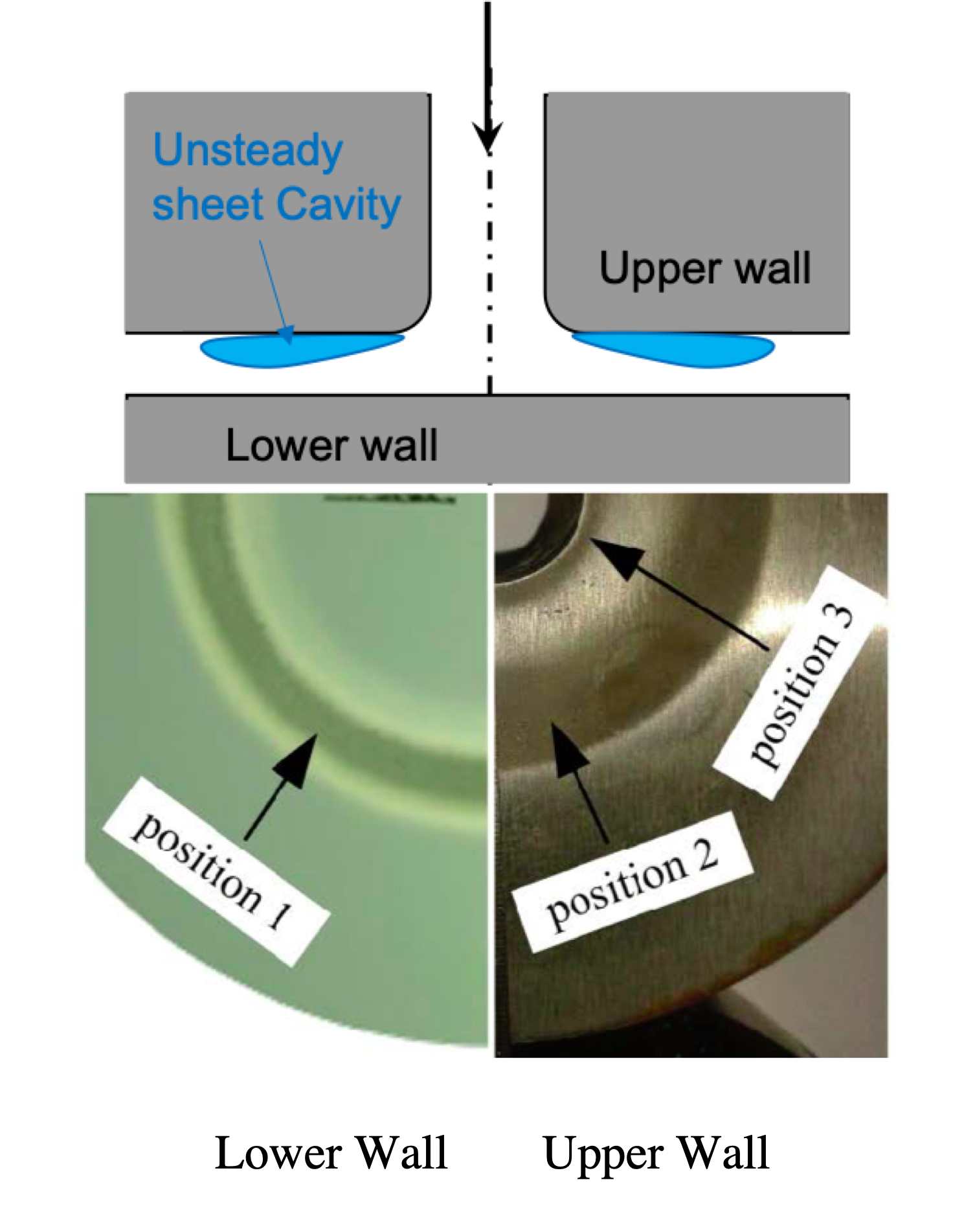}
\caption{}
\label{Fig:ExpErosionAxisNozzle}
\end{subfigure}
\begin{subfigure}{0.59\textwidth}
\centering
\includegraphics[width=\linewidth]{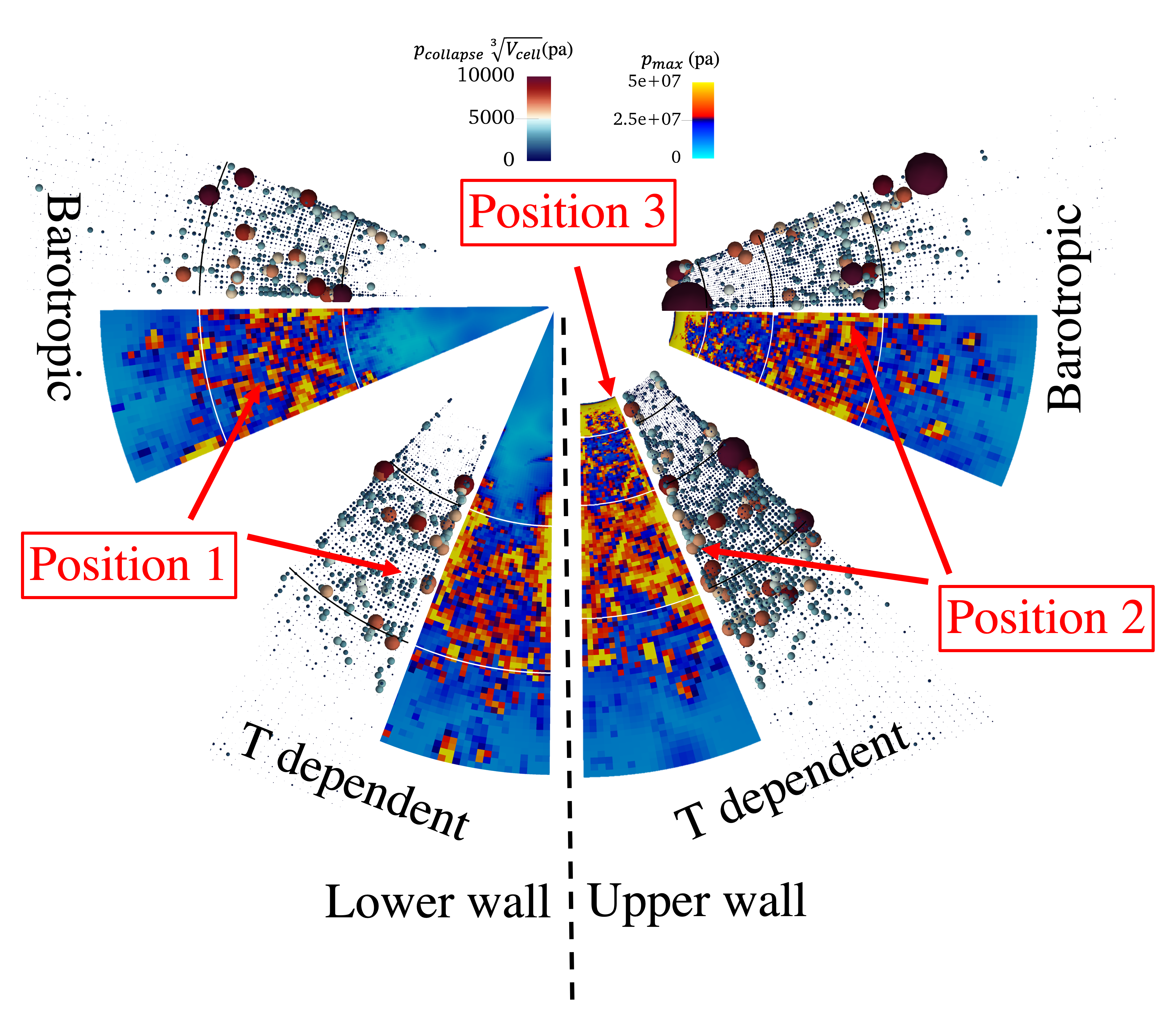}
\caption{}
\label{Fig:NumErosionAxisNozzle}
\end{subfigure}
\caption{Numerical and the experimental erosion pattern, (a) the erosion pattern in the experiment by \citet{franc2011impact} and (b) the predicted areas with high risk of cavitation erosion in the simulation with different mesh resolution.}
\label{Fig:AxisNozzleCom}
\end{figure}

As mentioned in the introduction section, one of the main advantages of the implemented solver is that the collapse-induced shock-waves captured by the solver can be analyzed to identify areas with a high risk of cavitation erosion. This analysis can be done based on the output of  \textbf{\texttt{erosionAssessment}} functionObject. To validate the implementation of this functionObject, Figure \ref{Fig:AxisNozzleCom} compares the output of this functionObject with the location of erosion in the experiments by \citet{franc2011impact} and \citet{gavaises2015visualisation}. In these experiments, erosion on the lower wall can be seen in the radius between  $r=19\,  \mathrm{mm}$ and  $r=32\,  \mathrm{mm}$ (position 1 in Figure \ref{Fig:ExpErosionAxisNozzle}), while on the upper wall, the eroded areas are located mostly in the radius between $r=17\,  \mathrm{mm}$ and  $r=27\ \mathrm{mm}$ (position 2 in Figure \ref{Fig:ExpErosionAxisNozzle}) and the radius smaller than  $r=11\,  \mathrm{mm}$.  Figure \ref{Fig:NumErosionAxisNozzle} presents the collapse locations near the lower and upper walls as well as the maximum pressure on these walls during simulations  which are obtained using the \textbf{\texttt{erosionAssessment}} functionObject.  In these figures, the location of the eroded areas is shown by black and white lines. It can be seen that on the lower wall, the distribution of the maximum pressure on the surface in the simulations with different cavitation models are quite similar and in a good agreement with the position 1 of the eroded areas on the lower wall in the experiment. The same good agreement can be seen for the detected collapse locations using the implemented functionObject in both simulations. For the upper wall, the location of the collapse events and the distribution of maximum pressure in position 3 agrees well with the experimental eroded areas in this region. In position 2, however, the distribution of the high maximum pressure and the location of aggressive collapses have a larger radial extension compared to the eroded areas in the experiment. This overprediction of the radial extension of position 2 is slightly more significant in the simulation with the temperature-dependent cavitation model compared to the one with the barotropic cavitation model.  It should be mentioned that the cavitation erosion shown in Figure \ref{Fig:ExpErosionAxisNozzle} is due to the interaction between the high mechanical load due to the collapse of cavitating structures and material. In the numerical results, however, only the strength of the collapse-induced shock-wave is used to identify areas with a high risk of cavitation erosion without considering the response of material which can explain the slight difference between the numerical results and experimental erosion pattern.  

\subsection{Cavitating flows in a micro-channel}
\begin{figure*}[h]
    \begin{minipage}[ht]{0.45\textwidth}
    \centering
     \includegraphics[width=\linewidth]{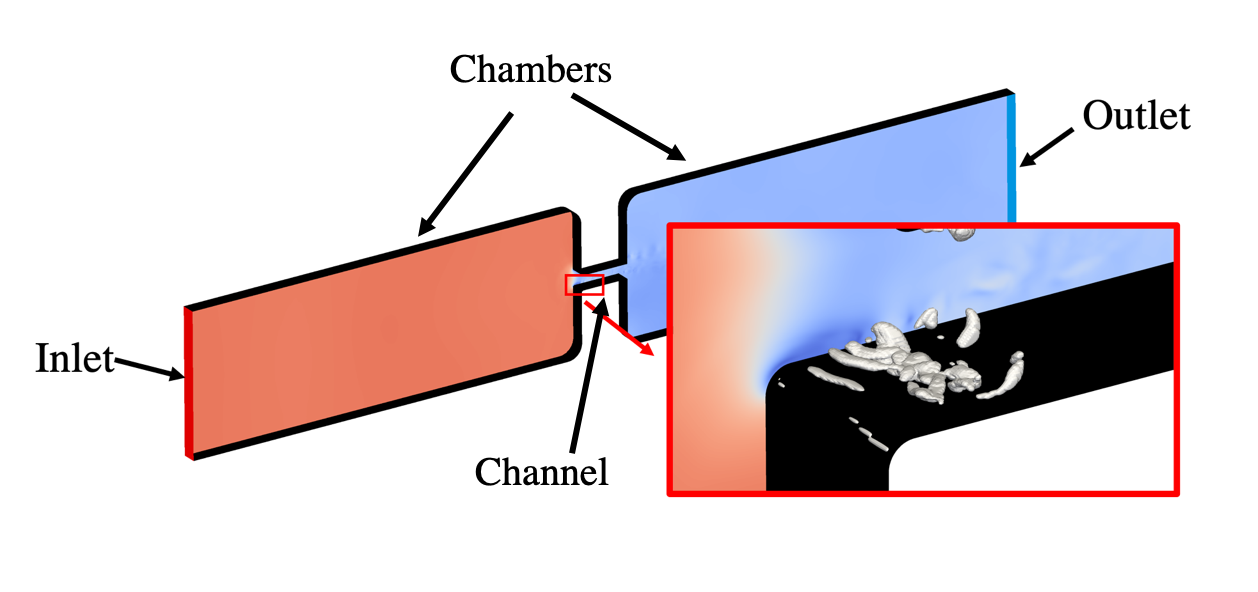}
    \captionof{figure}{Flow configuration of the cavitating flows in a micro-channel}
    \label{MicroChannelConfig}
  \end{minipage}
  \hspace{0.5cm}
  \begin{minipage}[hb]{0.4\textwidth}
        \captionof{table}{The studied flow condition for the cavitating flows in a micro-channel}
        \vspace{0cm}
        \label{MicroChannelStudiedConditions}
    \centering
\begin{tabular}{c|c|c|c}
Conditions & \multicolumn{1}{l|}{$\dot{m}_{inlet} (kg/s)$} & $U_{intlet}(\textup{m/s})$ & $\sigma$ \\ \hline
C1         & 0.01450                                       & 19.65                      & 0.54     \\ \hline
C2         & 0.01508                                       & 20.43                      & 0.20    
\end{tabular}
    \end{minipage}

 \end{figure*}

The last test case is the viscous simulations of cavitating flows in a throttle which resembles the flow configuration in the numerical study by \citet{egerer2014large}. In these simulations, Large Eddy Simulation (LES) with the WALE subgrid-scale model \citep{nicoud1999subgrid} is used for turbulence modeling and the barotropic cavitation model is used as the cavitation model. The flow configuration which is shown in Figure \ref{MicroChannelConfig}, includes two chambers connected by a narrow channel. The flow enters one of the chambers, passes through the narrow channel, and then is discharged to the other channel.  Due to the flow contraction at the entrance of the channel, the velocity increases, leading to a pressure drop which in turn results in the formation of cavitating structures in the channel, as shown in Figure \ref{MicroChannelConfig}. Two flow conditions are studied which are shown in Table \ref{MicroChannelStudiedConditions}. These conditions are the same as the ones in the numerical studies by \citet{egerer2014large} in order to be able to compare the results in this paper with these authors' numerical results. 

 \begin{figure}[H]
\includegraphics[width=\linewidth]{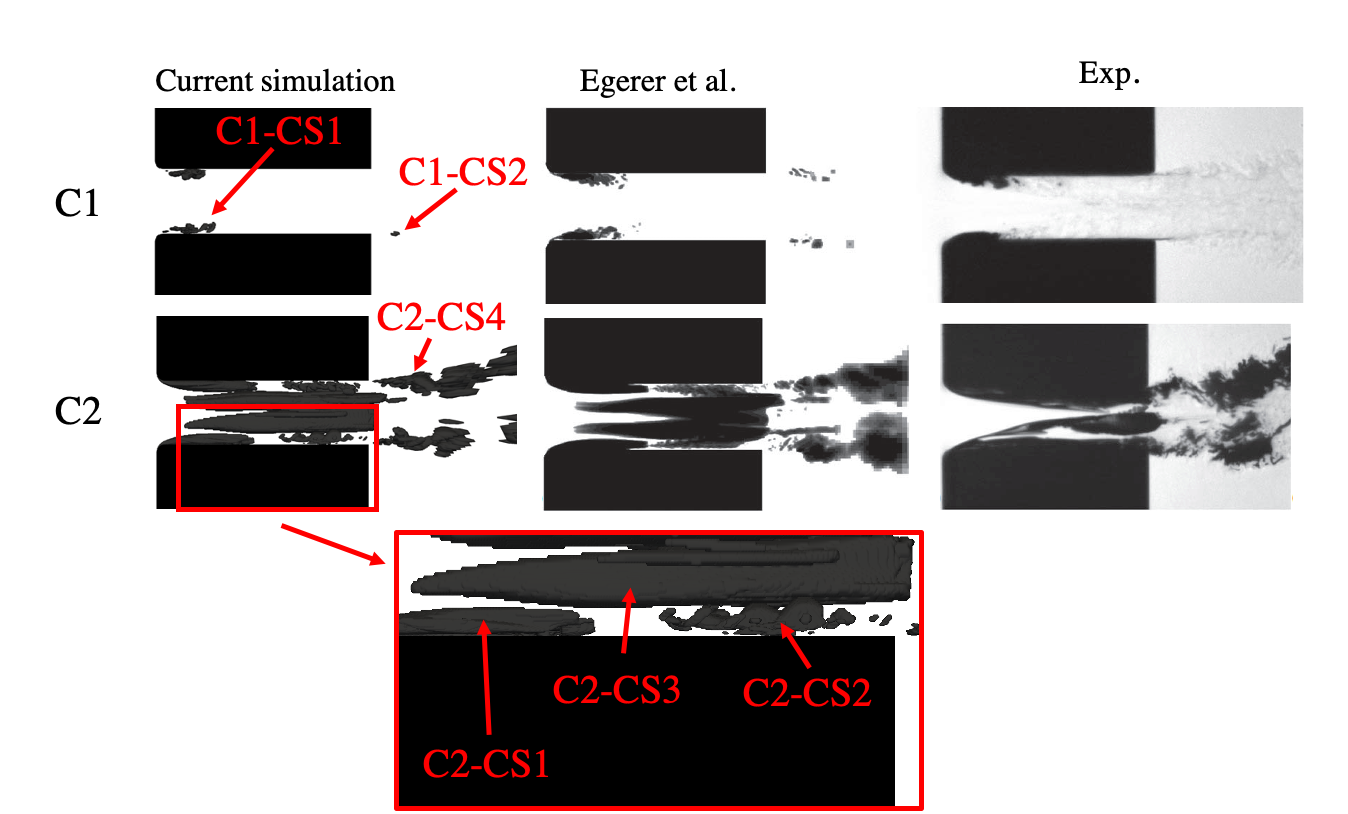}
\caption{Comparison between the cavitating regions in the numerical results by the implemented solver, the numerical results by \citet{egerer2014large}, and the experiment by \cite{iben2011laser}. }
\label{Comp_CavDyn_MicroChannel} 
\end{figure}

Figure \ref{Comp_CavDyn_MicroChannel} shows the cavitating structures captured by the implemented solver for the conditions C1 and C2. For comparison, the cavitating structures in the numerical study by \citet{egerer2014large} and experimental study by \citet{iben2011laser} are also shown in this figure.   The comparison made in these figures shows that the implemented solver can capture similar cavitating structures compared to ones in the reference simulation and experiment. In the condition C1, small cavitating structures, marked by C1-CS1 in Figure \ref{Comp_CavDyn_MicroChannel}, forms at the entrance of the channel on the lower and upper walls. According to \citet{egerer2014large}, these cavitating structures show periodic behavior with the frequency between 250-300 $\textup{kHz}$. In the simulation obtained by the implemented solver, similar period behavior has been observed; however, the shedding frequency is between 190-470 $\textup{kHz}$. It should be mentioned that the mesh resolution used in the simulation presented here is coarser than the one in the study by \citet{egerer2014large} which can explain the difference between the range of shedding frequencies. The images for the condition C1 also show that small cavitating structures, marked by C1-CS2, are formed in the shear layer between the flow exiting the channel and the flow in the chamber. In the condition C2, a stationary sheet cavity is formed on the upper and lower walls which is marked by C2-CS1. In addition to this sheet cavity, two different types of cavitating structures are formed in the channel (marked by C2-CS2 and CS3 in Figure \ref{Comp_CavDyn_MicroChannel}), which are detached from upper and lower walls. The structure C2-CS2 forms in the wake of the sheet cavity while the structure C2-CS3 starts from a region above the sheet cavity and is stretched up to the channel exit. Similar to the condition C1, it can be seen that in the condition C2, the cavitating structures are formed in the shear layer connected to the exit of the channel. 

\begin{figure}[ht!]
\begin{subfigure}{0.4\textwidth}
\centering
\includegraphics[width=\linewidth]{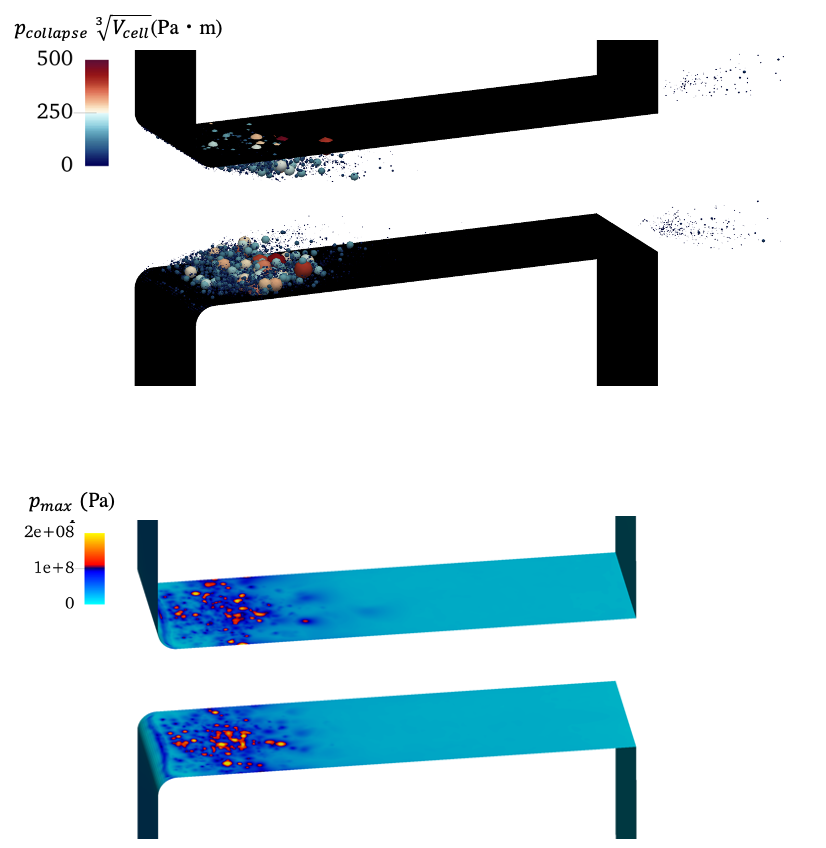}
\caption{Condition C1 in present simulation}
\label{Fig:C1_CurrentSim}
\end{subfigure}
\begin{subfigure}{0.4\textwidth}
\centering
\includegraphics[width=\linewidth]{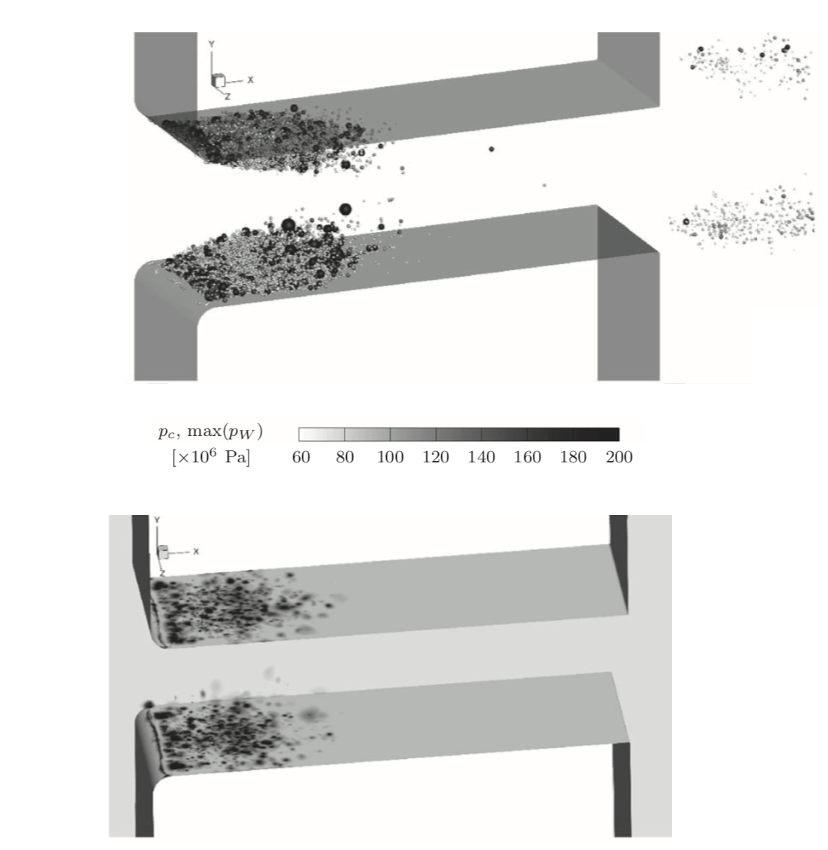}
\caption{Condition C1 in the study by \citet{egerer2014large}}
\label{Fig:C1_Egerer}
\end{subfigure}
\begin{subfigure}{0.4\textwidth}
\centering
\includegraphics[width=\linewidth]{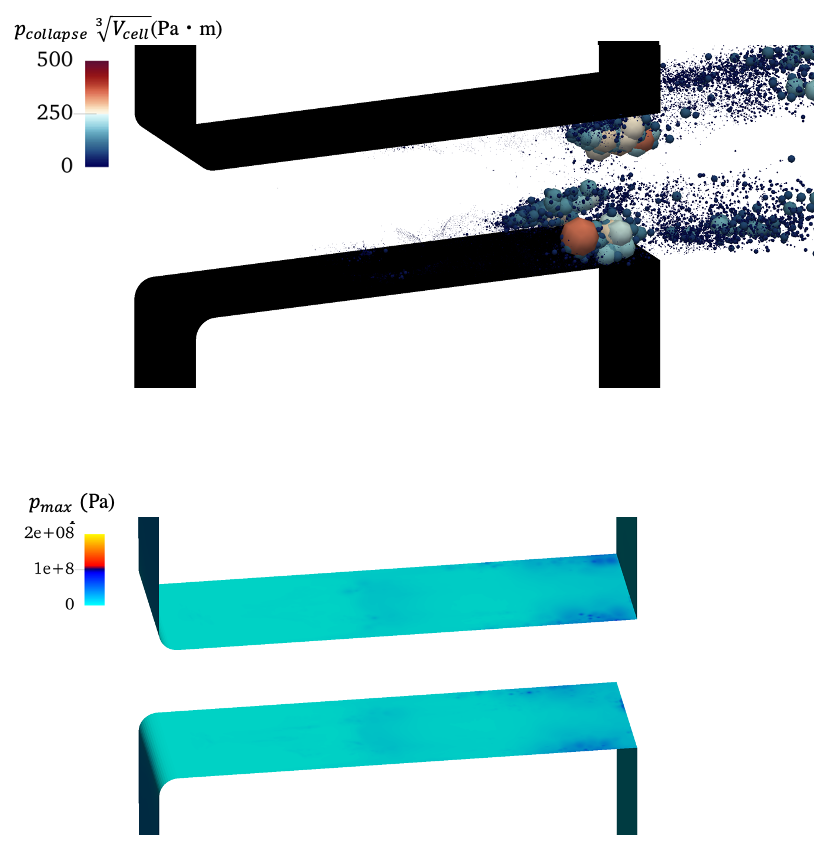}
\caption{Condition C2 in present simulation}
\label{Fig:C2_Current}
\end{subfigure}
\centering
\begin{subfigure}{0.4\textwidth}
\centering
\includegraphics[width=\linewidth]{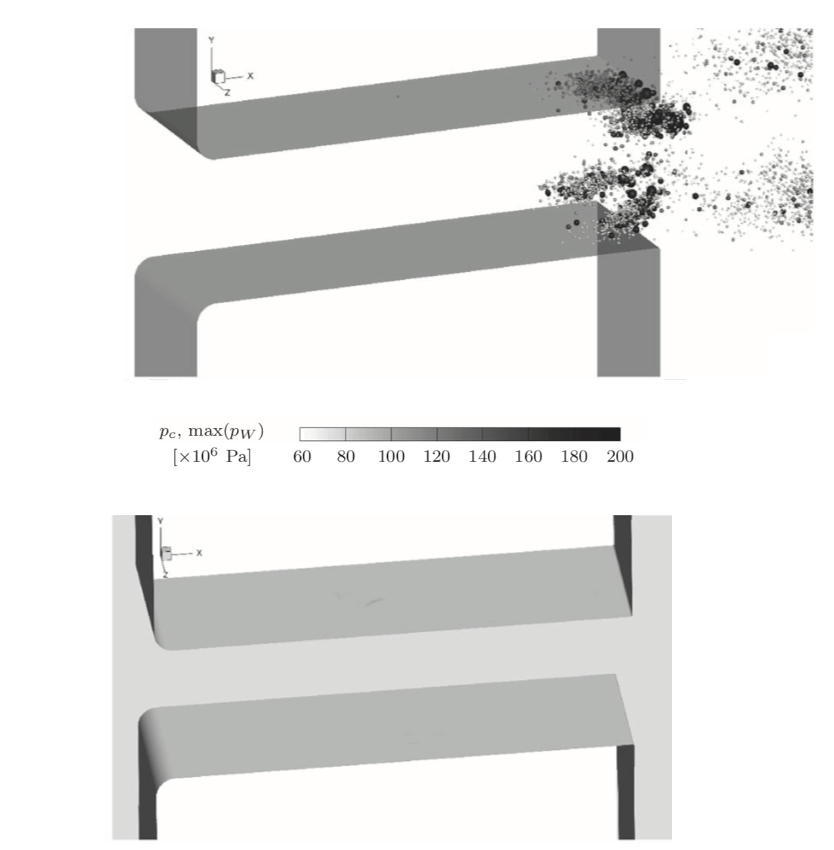}
\caption{Condition C2 in the study by \citet{egerer2014large}}
\label{Fig:C2_Egerer}
\end{subfigure}
\caption{Comparison between the erosion assessment performed by the \textbf{\texttt{erosionAssessment}} functionObject  and erosion assessment present in the study by \citet{egerer2014large}.}
\label{Fig:erosionAssessmentMicroChannel}
\end{figure}

Figure \ref{Fig:erosionAssessmentMicroChannel} shows the erosion assessment performed by the \textbf{\texttt{erosionAssessment}} functionObject  for the conditions C1 and C2. This figure also includes the erosion assessment in the study by \citet{egerer2014large}  for comparison. Both erosion assessments indicate that the flow in condition C1 has a higher risk of cavitation erosion compared to the condition C2. In the condition C1 shown in Figures \ref{Fig:C1_CurrentSim} and \ref{Fig:C1_Egerer}, collapse events can be seen at the entrance of the channel where unsteady cavitating structures C1-CS1 are formed according to Figure \ref{Comp_CavDyn_MicroChannel}. These collapse events are close to the surface and can induce high pressure on the surface as it can be seen in the distribution of maximum pressure shown in Figure \ref{Fig:C1_CurrentSim} and \ref{Fig:C1_Egerer}. The figures for the condition C1 also show some collapse events outside of the channel which are due to the cavitating structures formed in the shear layer at the exit of the channel (structures C1-CS2 in Figure \ref{Comp_CavDyn_MicroChannel}). In the condition C2, collapse events are mostly located at the exit of the channel. These collapse events are due to unsteady behavior in the downstream of the structures C2-CS2 and C2-CS3 in Figure \ref{Comp_CavDyn_MicroChannel} and the structures C2-CS4 in Figure \ref{Comp_CavDyn_MicroChannel} formed in the shear layer outside of the channel. The distribution of the maximum pressure on the surface indicates that these collapse events in the condition C2 do not produce high pressure on the surface due to their distance to the nearby surface.

\section{Conclusions}
In this paper, the implementation of a new density-based compressible solver for the simulation of cavitating flows, called \textbf{\texttt{dbnsCavitatingFoam}}, in the OpenFOAM framework is presented. This implementation which follows the OpenFOAM's programming standard is made publicly available for the first time and can serve as the platform for further development of similar solvers for cavitating flows. In the solver, the phase transition between liquid and vapor is modeled using thermodynamic equilibrium cavitation models. Two of these cavitation models, the temperature-dependent cavitation model and barotropic cavitation model by \citet{egerer2014large}, are implemented in a library, called \textbf{\texttt{equilibriumCavitationModels}},  which is used in the solver. The implemented solver has two other libraries,  \textbf{\texttt{dbns}} and \textbf{\texttt{dbnsCavTurbulenceModels}}. The \textbf{\texttt{dbns}} library includes the implementation of the Mach consistent numerical flux developed by \citet{schmidt2015low} while the \textbf{\texttt{dbnsCavTurbulenceModels}} library creates a new set of compressible turbulence models based on implemented models in OpenFOAM framework. In the \textbf{\texttt{dbnsCavitatingFoam}} solver, the compressibility of vapor and liquid phase is taken into account which means that collapse-induced shock-waves can be captured. As these shock-waves are one of the mechanisms of cavitation erosion, a post-processing tool, called \textbf{\texttt{erosionAssessment}}, is implemented which analyzes the strength of these shock-waves based on the work of \citet{mihatsch2015cavitation} and identifies the areas with a high risk of cavitation erosion.

For validation of the solver and to show its capabilities, the simulation of four test cases are presented.  These cases are a one-dimensional cavitating flow, a three-dimensional collapsing bubble, a cavitating flow in an axisymmetric nozzle, and cavitating flows in a micro-channel. In the first two simple cases, the numerical results are compared with the numerical simulation by \citet{sezal2009compressible} and analytical solution of Rayleigh equation \citep{rayleigh1917pressure} to check the implementation of the two cavitation models. In the third test case where the inviscid simulation of a cavitating flow in an axisymmetric nozzle is performed, the cavitating dynamics and the areas predicted with a high risk of cavitation erosion captured by the solver are compared with experimental studies by \citet{gavaises2015visualisation} and \citet{franc2011impact}.  This comparison shows that the captured cavitation dynamics and regions prone to high risk of cavitation erosion agree qualitatively with these experimental studies. The last case is LES simulation of cavitating flows in a micro-channel for two flow conditions. According to the numerical study by \citet{egerer2014large}, these two conditions have different risk of cavitation erosion. It is shown that this different risk of cavitation erosion can be captured by the implemented solver. Based on the location of aggressive collapse events obtained by  \textbf{\texttt{erosionAssessment}} post-processing tool, it can be concluded that in the condition with a lower risk of cavitation erosion, the distance between aggressive collapse events and the surface is larger compared to the condition with a lower risk of cavitation erosion. Due to this larger distance, shock-waves produced upon the aggressive collapse events in the condition with lower erosion risk are not able to produce high pressure on the surface, leading to a lower risk of cavitation in this condition. 

\section*{\small Acknowledgements}
This work is funded through the EU H2020 project CaFE, a Marie Sk{\l}odowska-Curie Action Innovative Training Network project, grant number 642536. The computations were performed on resources at Chalmers Centre for Computational Sciences and Engineering (C3SE) provided by the Swedish National Infrastructure for Computing (SNIC).

\bibliographystyle{plainnat}
\bibliography{Manuscripts}

\end{document}